\documentclass[twocolumn,epjc3]{svjour3}

\usepackage[english]{babel}

\usepackage{graphicx}
\usepackage{mathptmx}

\usepackage[numbers,sort&compress]{natbib}
\usepackage[caption=false]{subfig}

\usepackage{booktabs}
\usepackage{amsmath,amssymb}
\usepackage{slashed}

\makeatletter
\providecommand*{\diff}%
  {\@ifnextchar^{\DIfF}{\DIfF^{}}}
\def\DIfF^#1{%
  \mathop{\mathrm{\mathstrut d}}%
    \nolimits^{#1}\gobblespace}
\def\gobblespace{%
    \futurelet\diffarg\opspace}
\def\opspace{%
    \let\DiffSpace\!%
    \ifx\diffarg(%
      \let\DiffSpace\relax
     \else
      \ifx\diffarg[%
	\let\DiffSpace\relax
      \else
	\ifx\diffarg\{%
	  \let\DiffSpace\relax
	\fi\fi\fi\DiffSpace}

\usepackage[breaklinks,colorlinks,citecolor=blue,urlcolor=blue,linkcolor=blue]{hyperref}
\hypersetup{pdftitle=Radiative corrections for the decay \textbackslash Sigma\^{}0 -> \textbackslash Lambda e\^{}+ e\^{}-}

\journalname{Eur. Phys. J. C}
\def\makeheadbox{}

\allowdisplaybreaks

\begin{document}

\title{Radiative corrections for the decay $\Sigma^0\to\Lambda e^+e^-$}

\author{%
Tom\'{a}\v{s} Husek\thanksref{e1,addr1}
\and
Stefan Leupold\thanksref{e2,addr2}
}

\thankstext{e1}{e-mail: thusek@ific.uv.es}
\thankstext{e2}{e-mail: stefan.leupold@physics.uu.se}

\institute{%
IFIC, Universitat de Val\`encia -- CSIC, Apt.\ Correus 22085, E-46071 Val\`encia, Spain \label{addr1}
\and
Institutionen f\"or fysik och astronomi, Uppsala universitet, Box 516, S-75120 Uppsala, Sweden \label{addr2}
}

\date{}

\maketitle
\thispagestyle{plain}

\begin{abstract}
Electromagnetic form factors serve to explore the intrinsic structure of nucleons and their strangeness partners.
With electron scattering at low energies the electromagnetic moments and radii of nucleons can be deduced.
The corresponding experiments for hyperons are limited because of the unstable nature of the hyperons.
Only for one process this turns to an advantage: the decay of the neutral Sigma hyperon to a Lambda hyperon and a real or virtual photon.
Due to limited phase space the effects caused by the Sigma-to-Lambda transition form factors compete with the QED radiative corrections for the decay $\Sigma^0\to\Lambda e^+e^-$.
These QED corrections are addressed in the present work, evaluated beyond the soft-photon approximation, i.e., over the whole range of the Dalitz plot and with no restrictions on the energy of the radiative photon.
\end{abstract}

\section{Introduction}
\label{sec:intro}

Within the Standard Model of Particle Physics the elementary building blocks of matter have been identified as leptons and quarks.
Yet, it is still a challenge of contemporary subatomic physics to understand even the first level of compositeness, the structure of the nucleons as built from quarks.
Historically, strangeness and electromagnetic probes provided first clues about the intrinsic structure of nucleons and hadrons in general.
From SU(2) symmetry, i.e.\ isospin, alone it would have been hardly possible to isolate quarks as members of the fundamental multiplet.
Only by including strangeness and the approximate SU(3) flavor symmetry the emerging multiplet structure of observed hadrons allowed to propose the existence of the elementary quark triplet \cite{GellMann:1964nj}.
Even earlier, the substantial deviation of the magnetic $g$-factor of the proton from the value of 2, as proposed by Dirac's theory, provided another hint that the proton cannot be as elementary as the electron \cite{1933ZPhy...85....4F}.
Subsequently, the corresponding analysis was extended to the scattering of electrons on nucleons to reveal more of the intrinsic structure of the nucleon \cite{Kirk:1972xm,Kendall:1991np}.
The nucleon spin crisis \cite{Aidala:2012mv} and the puzzle about the charge radius of the proton \cite{Carlson:2015jba} demonstrate that our understanding of the structure of the proton and the neutron is still very limited.

In this situation a close collaboration between theory and experiment is instrumental in making significant progress.
Concerning the electromagnetic form factors of the nucleon we refer to the reviews \cite{Denig:2012by,Punjabi:2015bba} and references therein.
A general attitude, when it comes to the study of a complex system, is to ask the question, what happens to the system if parts of their components are slightly modified.
For electromagnetic form factors the natural modifications are a spin-flip and/or a flavor change.
This extends the form factor business from nucleons to hyperons \cite{Aubert:2007uf,Faldt:2013gka,Dobbs:2014ifa,Faldt:2016qee,Faldt:2017kgy,Perotti:2018wxm,Ablikim:2019vaj,Faldt:2019zdl} and to transition form factors between spin 1/2 and 3/2 states \cite{Pascalutsa:2006up}.

The present work should be seen as part of an endeavor to motivate and assist experimental activities aiming at the extraction 
of electromagnetic hyperon (transition) form factors.
As already pointed out, we expect that the study of hyperons will provide an additional angle to look at the structure of nucleons, complementary to the elastic nucleon form factors and the nucleon-to-Delta transition form factors.
Hyperons come with additional challenges, but also opportunities:
Since hyperons are unstable, the experimental opportunities to collide hyperons with electrons are technically very limited.
This moves the focus from the space-like to the time-like region concerning the virtuality of the photon.
For elastic form factors (in a generalized sense invoking crossing symmetry) the experimentally accessible region of photon virtuality starts at twice the hyperon mass, i.e.\ one has to study the reactions $e^+e^-$ to a hyperon--antihyperon pair.
For transition form factors, however, there is a low-energy window that allows for access to the electromagnetic radii, i.e.\ to the slopes of the form factors at vanishing photon virtuality.
Such quantities enter the Dalitz decays of a hyperon into another hyperon plus an electron--positron pair.

Transitions from the spin 3/2 decuplet states to the spin 1/2 octet members are addressed elsewhere \cite{Junker:2019vvy}.
In the present work we will focus on the one possible electromagnetic transition within the ground-state octet, namely the Dalitz decay $\Sigma^0\to\Lambda e^+e^-$.
Of course, the electric and magnetic transition form factors enter the decay rate.
However, the phase space for the lepton pair is rather limited, $M_{\Sigma^0}-M_\Lambda\approx77\,$MeV \cite{pdg}.
Consequently, even the differential decay rate is dominated by the transition magnetic moment, which can and has been determined from the simpler two-body decay $\Sigma^0\to\Lambda\gamma$ \cite{pdg}.
As can be expected and as we will see below, the numerical impact of the form factors themselves, i.e.\ essentially from the electric and magnetic transition radii, is rather limited.
As it turns out, the impact from the hadronic structure competes with the radiative corrections imposed by Quantum Electrodynamics (QED).
The transition form factors themselves are addressed in a complementary work \cite{Granados:2017cib} while the present work is devoted to these QED corrections.

From the experimental point of view the transition of the $\Sigma^0$ to a $\Lambda$ and a {\em real} photon has been measured \cite{pdg}.
The Dalitz decay $\Sigma^0\to\Lambda e^+e^-$ has not been observed yet.
The value for its branching ratio quoted in Ref.~\cite{pdg} is purely based on a theoretical calculation \cite{Feinberg:1958zz} that neglects hadronic structure effects and QED corrections.
Nonetheless, it can be expected that this value for the branching ratio will be fairly accurate.
To reveal deviations from leading-order QED and from a point-like hadron structure requires differential data for this Dalitz decay, i.e.\ data with high statistics and high precision.
On the other hand, hyperons move more and more in the focus of experimental activities.
High-energy time-like form factors of hyperons have recently been addressed by BaBar \cite{Aubert:2007uf}, CLEO-c \cite{Dobbs:2014ifa} and BES-III \cite{Ablikim:2019vaj}.
With the advent of the planned Facility for Antiproton and Ion Research (FAIR) \cite{fair-homepage} a hyperon factory will start to operate.
Both in the proton-antiproton collisions studied by PANDA \cite{Lutz:2009ff} and in the proton-proton collisions studied by HADES \cite{Ramstein:2019kaz} hyperons will be copiously produced and there are detectors dedicated to studying the hyperons and their properties \cite{Lutz:2009ff,Agakishiev:2014kdy}.
Thus it can be expected that it will be possible in the future to collect enough data for an experimental determination of not only the branching ratio but also the differential Dalitz decay width of $\Sigma^0\to\Lambda e^+e^-$.
The present work will serve to disentangle the QED-correction effects from the effects caused by the intrinsic structure of the hyperons.
In turn, information about the intrinsic structure of the hyperons will provide a new angle on the structure of the nucleon.

Radiative corrections to the $\Sigma^0\to\Lambda e^+e^-$ process were already studied by Sidhu and Smith~\cite{Sidhu:1972rx}.
In that work, the corrections to the decay rate as well as to the differential decay width were calculated.
In the latter case, the soft-photon approximation was used.
Moreover, the one-photon-irreducible (1$\gamma$IR) contribution was not calculated, based --- apart from its apparent difficulty --- on the assumption that it was negligible.
Similarly, it was argued that the correction to the $\Sigma^0\Lambda\gamma$ vertex did not affect the measurement of the slope of the form factor and thus was irrelevant for the scope of Ref.~\cite{Sidhu:1972rx} and further left out from the discussion.

In the present paper we decided to reinvestigate these claims and calculate the contribution of (nearly) all the QED diagrams at next-to-leading order (NLO) explicitly.
Most importantly, we present the bremsstrahlung contribution (regarding the lepton legs) {\em beyond} the soft-photon approximation, i.e.\ including the hard-photon corrections.
Our result represents the complete {\em inclusive} QED radiative correction at NLO, leaving out only the bremsstrahlung correction related to the hyperon legs.
Here, we agree with Ref.~\cite{Sidhu:1972rx} that it is safe to neglect this contribution:
Firstly, because of the significantly higher rest mass of the hyperons compared to the mass of the electron (related to which the bremsstrahlung indeed represents a significant contribution).
Secondly, since $\Sigma^0$ and $\Lambda$ are neutral, the magnetic moment dominates the photon emission, which leads to an additional suppression.
In particular, no infrared (IR) divergent terms are present to enhance the effect of this contribution.

Our motivation to calculate the 1$\gamma$IR contribution explicitly (and for completeness also the correction to the $\Sigma^0\Lambda\gamma$ vertex) is based on the fact that it already happened in the literature that the assumption that a particular contribution was negligible turned out to be incorrect.
For instance, the 1$\gamma$IR contribution to the radiative corrections for the neutral-pion Dalitz decay $\pi^0\to e^+e^-\gamma$ was, due to inappropriate assumptions and arguments based on Low's theorem~\cite{Low:1958sn,Adler:1966gc,Pestleau:1967snm}, considered negligible
and left out from the classical work~\cite{Mikaelian:1972yg}; see also Ref.~\cite{Lambin:1985sb}.
The exact calculation, in contrary, shows its significance~\cite{Tupper:1983uw,Tupper:1986yk,Kampf:2005tz,Husek:2015sma}.
The impact on the form factor slope is considerable, especially in view of a precision measurement or calculation~\cite{Husek:2018qdx}.
Similarly, it can happen that an approximate formula is derived which, however, differs significantly from the exact calculation when carried out:
Accidental cancellations or loop enhancements might take place.
As an example, let us mention the approximate calculation of the two-loop virtual radiative corrections to the neutral-pion rare decay $\pi^0\to e^+e^-$~\cite{Bergstrom:1982wk,Dorokhov:2008qn} and their significant difference to the exact calculation~\cite{Vasko:2011pi}, which seems to be one of the main sources of the theory--experiment discrepancy~\cite{Dorokhov:2007bd,Husek:2014tna,Husek:2015wta}.

The rest of the paper is structured in the following way.
In Section~\ref{sec:LO} we fix our notation and conventions and present the leading-order (LO) results.
In the subsequent sections we then discuss individual radiative corrections.
In Section~\ref{sec:virt} we present the results for the virtual corrections related to photon and lepton legs.
We discuss the corresponding bremsstrahlung correction in Section~\ref{sec:BS} and show a compact approximate (although numerically satisfactory) result there.
Additional related expressions are provided in \ref{app:J}.
The 1$\gamma$IR correction is treated in Section~\ref{sec:1gIR}, which is further complemented by additional four appendices.
In Section~\ref{sec:SLg} we discuss the (virtual) QED correction to the $\Sigma^0\Lambda\gamma$ vertex.
We conclude with results and discussion in Section~\ref{sec:res}.

\section{Definitions and the leading order}
\label{sec:LO}

In what follows we briefly introduce the notation.
We denote the four-momenta of the neutral Sigma baryon (of mass $M_\Sigma$), Lambda hyperon (of mass $M_\Lambda$), electron (of mass $m$) and positron by $p_1$, $p_2$, $q_1$ and $q_2$, respectively.
Thus it holds $p_1=p_2+q_1+q_2$, provided bremsstrahlung is not included.

As for the Lorentz structure of the $\Sigma^0\Lambda\gamma$ vertex we write \cite{Kubis:2000aa,Granados:2017cib}
\begin{equation}
\langle0|j^\mu|\Sigma^0\bar\Lambda\rangle
=e\bar v_\Lambda(\vec p_2)G^\mu(p_1+p_2)u_\Sigma(\vec p_1)\,,
\label{eq:SL_Lorentz}
\end{equation}
with
\begin{equation}
G^\mu(q)
\equiv
\bigg[\gamma^\mu-\Delta_M\frac{q^\mu}{q^2}\bigg]G_1\big(q^2\big)
-\frac{i\sigma^{\mu\nu}q_\nu}{2\hat M}G_2\big(q^2\big)\,,
\label{eq:Gon}
\end{equation}
where we defined (for the outgoing four-momentum $q$) the Dirac and Pauli transition form factors as $G_1$ and $G_2$, respectively.
Above, we used $\sigma^{\mu\nu}\equiv\frac i2[\gamma^\mu,\gamma^\nu]$, $\hat M\equiv(M_\Sigma+M_\Lambda)/2$ and $\Delta_M\equiv (M_\Sigma-M_\Lambda)$.
For real photons ($q^2=0$), the transition form factors become $G_1(0)=0$ and $G_2(0)=\kappa$.
Here $\kappa \approx 1.98$ is related to the transition magnetic moment \cite{pdg} $\mu = \kappa e/(2 \hat M)$.

In an equivalent way we introduce in the case of the $e^+e^-\gamma$ vertex (again, $q$ is outgoing)%
\footnote{Using $q=q_1+q_2$, this notation is consistent with $$F^\mu(q_1,q_2)\equiv\gamma^\mu\big[F_1\big(q^2\big)+F_2\big(q^2\big)\big]
-\frac1{2m}(q_1-q_2)^\mu F_2\big(q^2\big)\,.$$}
\begin{equation}
F^\mu(q)
\equiv
\gamma^\mu F_1\big(q^2\big)
-\frac{i\sigma^{\mu\nu}q_\nu}{2m}F_2\big(q^2\big)\,,
\end{equation}
which comes into play when virtual radiative corrections are also considered.

The matrix element of the $\Sigma^0\to\Lambda e^+e^-$ process (for the one-photon-exchange topology) is then written in a simple form
\begin{equation}
\begin{split}
i\mathcal{M}(p_2,q_1,q_2)
&=\frac{-ig_{\mu\nu}}{(q_1+q_2)^2}\,
\bigl[-\tilde\Pi\big((q_1+q_2)^2\big)\bigr]\\
&\times[(ie)\bar u_\Lambda(\vec p_2) G^\mu(p_1-p_2) u_\Sigma(\vec p_1)]\\
&\times[(-ie)\bar u_e(\vec q_1) F^\nu(-q_1-q_2) v_e(\vec q_2)]\,,
\end{split}
\label{eq:M}
\end{equation}
where $\tilde\Pi(q^2)$ includes vacuum-polarization effects that we will eventually include.
Putting $\tilde\Pi(q^2)=-1$ corresponds to the photon propagator with no insertion.
Note that due to the conservation of the electromagnetic current, the part of Eq.~(\ref{eq:Gon}) proportional to $\Delta_MG_1(q^2)$ vanishes after the contraction with the leptonic part in Eq.~(\ref{eq:M}).

It becomes convenient to introduce a dimensionless variable
\begin{equation}
x
\equiv
\frac{(q_1+q_2)^2}{\Delta_M^2}\,,
\end{equation}
which stands for the normalized square of the total energy of the electron--positron pair in its center-of-mass system (CMS) (or simply of the electron--positron pair
invariant mass).
We also define the following small parameter:%
\footnote{Note that this definition of $\rho$ is different from the one used in Ref.~\cite{Sidhu:1972rx}.}
\begin{equation}
\rho
\equiv
\frac{(M_\Sigma-M_\Lambda)^2}{(M_\Sigma+M_\Lambda)^2}=\frac{\Delta_M^2}{4\hat M^2}\,.
\label{eq:rho}
\end{equation}
Numerically, $\rho\approx1.1\times10^{-3}\ll1$.
The second independent variable to describe the kinematics of the 3-body decay is chosen as
\begin{equation}
y
\equiv
-\frac{2p_1\cdot(q_1-q_2)}{\lambda^{\frac12}(p_1^2,p_2^2,(q_1+q_2)^2)}\,,
\label{eq:y}
\end{equation}
which has the meaning of a rescaled cosine of the angle between the directions of the incoming (decaying) hyperon and the (outgoing) positron in the electron--positron CMS.
The K\"all\'en triangle function, generally defined as
\begin{equation}
\lambda(a,b,c)
\equiv
a^2+b^2+c^2-2ab-2ac-2bc\,,
\end{equation}
reduces in the case used in Eq.~(\ref{eq:y}) to
\begin{equation}
\lambda(M_\Sigma^2,M_\Lambda^2,\Delta_M^2x)
=\rho(1-x)(1-\rho x)(2\hat M)^4
\equiv\lambda(x)\,.
\label{eq:lambdax}
\end{equation}
Finally, we introduce $\nu\equiv2m/\Delta_M$ and
\begin{equation}
\beta
\equiv\beta(x)
\equiv\sqrt{1-\frac{\nu^2}x}\,,
\end{equation}
so the limits on kinematic variables $x$ and $y$ are simply given by
\begin{equation}
x\in [\nu^2,1]\,,
\quad
y\in [-\beta,\beta]\,.
\end{equation}
Note that $\nu\approx1.3\times10^{-2}\ll1$ is beside $\rho$ another small parameter.

Next, we define the electric ($G_\text{E}$) and magnetic ($G_\text{M}$) form factors in the following manner:
\begin{equation}
\begin{split}
G_\text{E}(q^2)
&\equiv
G_1(q^2)+\frac{q^2}{4\hat M^2}G_2(q^2)\,,\\
G_\text{M}(q^2)
&\equiv
G_1(q^2)+G_2(q^2)\,,
\end{split}
\label{eq:GEM}
\end{equation}
which in turn means
\begin{equation}
\begin{split}
G_1(\Delta_M^2x)
&=\frac{G_\text{E}(\Delta_M^2x)-\rho x G_\text{M}(\Delta_M^2x)}{1-\rho x}\,,\\
G_2(\Delta_M^2x)
&=\frac{G_\text{M}(\Delta_M^2x)-G_\text{E}(\Delta_M^2x)}{1-\rho x}\,.
\end{split}
\label{eq:G12}
\end{equation}
The modulus square of the matrix element, summed over the spins of $\Lambda$, electron and positron, and averaged over the spins of $\Sigma^0$ is given by
\begin{equation}
\begin{split}
\overline{|\mathcal{M}(x,y)|^2}
=\frac{2e^4(1-x)}{\rho x^2}
\overline{|\mathcal{M}_G(x,y)|^2}\,.
\end{split}
\label{eq:M2}
\end{equation}

The advantage in the use of the electric and magnetic over the Dirac and Pauli transition form factors lies in the fact that 
the quantity \eqref{eq:M2} contains no interference terms of $G_\text{E}$ and $G_\text{M}$.
Therefore also the two-fold differential decay width
\begin{equation}
\frac{\diff^2\Gamma(x,y)}{\diff x\diff y}
=\frac1{2M_\Sigma}\frac{\Delta_M^2\lambda^{\frac12}(x)}{32(2\pi)^3M_\Sigma^2}\,\overline{|\mathcal{M}(x,y)|^2}
\label{eq:dGxy}
\end{equation}
can be expressed in a form orthogonal in $G_\text{E}(q^2)$ and $G_\text{M}(q^2)$.
Indeed, squaring the hadronic part (corresponding to the term in the second square brackets) of the matrix element (\ref{eq:M}) reveals
\begin{equation}
\begin{split}
&\mathcal{M}_{(\text{H})}^\mu(\mathcal{M}_{(\text{H})}^\nu)^*
=-2e^2\biggl[
\Delta_M^2(1-x)|G_\text{M}(\Delta_M^2x)|^2g^{\mu\nu}+\dots\biggr.\\
&+\biggl.\frac{\rho x |G_\text{M}(\Delta_M^2x)|^2-|G_\text{E}(\Delta_M^2x)|^2}{1-\rho x}(p_1+p_2)^\mu(p_1+p_2)^\nu
\biggr],
\end{split}
\end{equation}
where the dots stand for similar terms proportional to $(p_1-p_2)^\alpha$.
These terms are also purely quadratic in the form factors and vanish upon contraction with the leptonic part.

\begin{figure}[t]
\centering
\includegraphics[width=0.6\columnwidth]{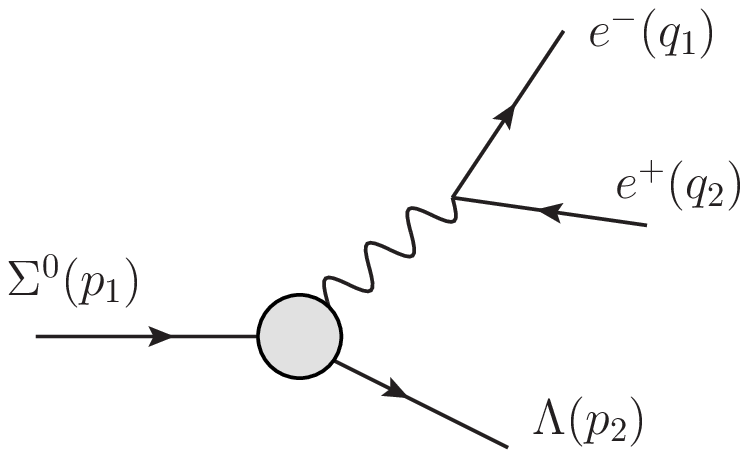}
\vspace{4mm}
\caption{
The leading-order diagram for the $\Sigma^0\to\Lambda e^+e^-$ process in the QED expansion.
The shaded blob corresponds to the $\Sigma^0\Lambda\gamma$ transition form factor.
}
\label{fig:LO}
\end{figure}

For the LO contribution for the $\Sigma^0\to\Lambda e^+e^-$ process --- which means putting simply $F_1(\Delta_M^2x)=1$, $F_2(\Delta_M^2x)=0$ and $\tilde\Pi(\Delta_M^2x)=-1$ in Eq.~(\ref{eq:M}) --- the matrix element can be written as
\begin{equation}
\begin{split}
&i\mathcal{M}^\text{LO}
=\frac{i^3e^2}{\Delta_M^2x}\big[\bar u_\Lambda(\vec p_2)\gamma_\sigma u_\Sigma(\vec p_1)\big]\big[\bar u_e(\vec q_1)\gamma_\tau v_e(\vec q_2)\big]\\
&\times\left\{\big[G_1(\Delta_M^2x)+G_2(\Delta_M^2x)\big]g^{\sigma\tau}-{G_2(\Delta_M^2x)}\frac{p_1^\sigma p_2^\tau}{M_\Sigma\hat M}\right\}\,;
\end{split}
\label{eq:MLO}
\end{equation}
see Fig.~\ref{fig:LO} for the associated Feynman diagram.
Thus we obtain 
\begin{equation}
\begin{split}
&\overline{|\mathcal{M}_G^\text{LO}(x,y)|^2}\\
&=(1-y^2)|G_\text{E}(\Delta_M^2x)|^2+\rho x\bigg(1+y^2+\frac{\nu^2}{x}\bigg)|G_\text{M}(\Delta_M^2x)|^2\,.
\end{split}
\label{eq:MGLOxy}
\end{equation}

At this point we are ready to define our approximation scheme.
The dominant contribution to the Dalitz distribution \eqref{eq:dGxy} is obtained from the LO QED process and neglecting the intrinsic structure of the hyperons.
Technically this is achieved by neglecting the $q^2=\Delta_M^2x$ dependence of the transition form factors, i.e.\ $G_{\text{E}}(q^2)\to0$ and $G_{\text{M}}(q^2)\to\kappa$.
In this work we consider NLO QED corrections proportional to the fine-structure constant $\alpha$ and corrections to the hadronic form factors that are linear in the ratio $\rho$, defined in Eq.~\eqref{eq:rho}.
All this is relative to the dominant contribution.
We can write
\begin{equation}
\begin{split}
G_\text{M}(\Delta_M^2x)
&\simeq\kappa\left(1+\rho x\,4\hat M^2\frac16\langle r^2_{\text{M}}\rangle\right),\\
G_\text{E}(\Delta_M^2x)
&\simeq\rho x\,4\hat M^2\frac16\langle r^2_{\text{E}}\rangle\,,
\end{split}
\label{eq:expandGEGM}
\end{equation}
where we introduced the transition radii via
\begin{equation}
\begin{split}
\langle r^2_{\text{M}}\rangle
&\equiv\frac6{\kappa}\;\frac{\diff G_\text{M}(q^2)}{\diff q^2}\bigg|_{q^2=0}\,,\\
\langle r^2_{\text{E}}\rangle
&\equiv6\;\frac{\diff G_\text{E}(q^2)}{\diff q^2}\bigg|_{q^2=0}\,.
\end{split}
\end{equation}
Note that the magnetic part is multiplied by a factor of $\rho$ in Eq.~\eqref{eq:MGLOxy}.
Thus the dominant contribution there is linear in $\rho$ and the considered corrections are $\sim\rho^2$.
What is neglected are the $\rho^3$ contributions in Eq.~\eqref{eq:MGLOxy}.

To justify our approximation scheme further, we note that hadronic radii are at most of order 1 fm.
Thus $\langle r^2\rangle\le(1\,\text{fm})^2\approx25\,$GeV$^{-2}$.
As a consequence of this estimate, the combination $\frac23\rho\hat M^2\langle r^2\rangle$ in Eq.~\eqref{eq:expandGEGM} is smaller than $0.03$.
Thus it makes sense to keep in Eq.~\eqref{eq:expandGEGM} the corrections suppressed by $\rho$, which are numerically comparable to the considered QED corrections.
The corrections suppressed by $\rho^2$, on the other hand, can be safely ignored.

Finally we add a theoretical estimate \cite{Kubis:2000aa,Granados:2017cib}.
The electric radius is much smaller than the magnetic one, $\vert \langle r^2_{\text{E}} \rangle \vert \ll \langle r^2_{\text{M}} \rangle$.
As a consequence of these considerations, we could completely ignore the electric transition form factor but we should keep the magnetic one and the linearized version of its $q^2$ dependence.
In most of our calculations we stick to Eq.~\eqref{eq:expandGEGM}.
Whenever we neglect in addition $\langle r^2_{\text{E}} \rangle$, we will spell it out explicitly.

We can thus simply rewrite Eq.~(\ref{eq:MLO}) substituting $G_1(q^2)\to G_\text{M}(q^2)-G_2(q^2)$ (from Eq.~(\ref{eq:GEM})) and taking only $G_2(\Delta_M^2x)\simeq G_\text{M}(\Delta_M^2x)/(1-\rho x)$ (from Eq.~(\ref{eq:G12})).
Using in addition $\rho x\ll1$ (which consequently translates into $G_1(\Delta_M^2x)\simeq0$) and the Dirac equation, we obtain
\begin{equation}
\begin{split}
i\mathcal{M}^\text{LO}
&\simeq\frac{i^3e^2}{\Delta_M^2x}
\,G_\text{M}(\Delta_M^2x)\left(g^{\sigma\tau}
-\frac{p_1^\sigma p_2^\tau}{M_\Sigma\hat M}\right)\\
&\times\big[\bar u_\Lambda(\vec p_2)\gamma_\sigma u_\Sigma(\vec p_1)\big]
\big[\bar u_e(\vec q_1)\gamma_\tau v_e(\vec q_2)\big]\,.
\end{split}
\label{eq:MLOb}
\end{equation}
Hence we arrive at
\begin{equation}
\overline{|\mathcal{M}^\text{LO}(x,y)|^2}
\simeq
2e^4|G_\text{M}(\Delta_M^2x)|^2
\frac{(1-x)}{x}
\bigg(1+y^2+\frac{\nu^2}{x}\bigg)
\label{eq:MLO2}
\end{equation}
for the LO matrix element squared.
The two-fold differential decay rate at LO reads (inserting Eq.~(\ref{eq:MLO2}) into Eq.~(\ref{eq:dGxy}))
\begin{equation}
\begin{split}
&\frac{\diff^2\Gamma^\text{LO}(x,y)}{\diff x\diff y}\\
&\simeq\frac{\alpha^2\Delta_M^2\lambda^{\frac12}(x)}{16\pi M_\Sigma^3}\,|G_\text{M}(\Delta_M^2x)|^2
\frac{(1-x)}{x}
\left(1+y^2+\frac{\nu^2}{x}\right).
\end{split}
\label{eq:dLOxy}
\end{equation}
Integrating Eq.~(\ref{eq:dLOxy}) over $y$, we find the one-fold differential decay width
\begin{equation}
\begin{split}
&\frac{\diff\Gamma^\text{LO}(x)}{\diff x}\\
&\simeq\frac{\alpha^2\Delta_M^2\lambda^{\frac12}(x)}{16\pi M_\Sigma^3}\,|G_\text{M}(\Delta_M^2x)|^2
\frac{(1-x)}{x}
\frac{8\beta}{3}
\left(1+\frac{\nu^2}{2x}\right).
\end{split}
\end{equation}

Going beyond LO, it is convenient to introduce the NLO correction $\delta$ to the LO differential decay width, which allows us to write schematically $\diff\Gamma=(1+\delta+\dots)\,\diff\Gamma^\text{LO}$.
In particular, we define
\begin{equation}
\delta(x,y)
=\frac{\diff^2\Gamma^\text{NLO}}{\diff x\diff y}\bigg/\frac{\diff^2\Gamma^\text{LO}}{\diff x\diff y}\,,\quad
\delta(x)
=\frac{\diff\Gamma^\text{NLO}}{\diff x}\bigg/\frac{\diff\Gamma^\text{LO}}{\diff x}\,.
\label{eq:dxydx}
\end{equation}
Related to the work documented in Refs.~\cite{Husek:2015sma,Husek:2017vmo}, such a correction can be divided into the following parts emphasizing the respective origin:
\begin{equation}
\delta
=\delta^\text{virt}+\delta^\text{BS}+\delta^{1\gamma\text{IR}}+\delta_{\Sigma^0\Lambda\gamma}^\text{virt}\,.
\label{eq:delta_origins}
\end{equation}
Here, the superscript ``virt'' stands for the virtual radiative corrections, $\delta^\text{BS}$ for the bremsstrahlung and $\delta^{1\gamma\text{IR}}$ for the 1$\gamma$IR contribution.
In our approach, the latter is treated separately from $\delta^\text{virt}$ for reasons of historical development, complexity and topology.
The part of the virtual correction associated to the $\Sigma^0\Lambda\gamma$ vertex, $\delta_{\Sigma^0\Lambda\gamma}^\text{virt}$, is treated separately from the corrections $\delta^\text{virt}$ related to photon and lepton legs.
The associated bremsstrahlung from the baryon legs is entirely neglected throughout this work.

As a trivial consequence of previous equations, having knowledge of $\delta(x,y)$ allows for obtaining $\delta(x)$ using the following prescription:
\begin{equation}
\delta(x)
=\left(\frac{\diff\Gamma^\text{LO}(x)}{\diff x}\right)^{-1}
\int_{-\beta}^{\beta}\diff y\,
\delta(x,y)
\frac{\diff^2\Gamma^\text{LO}(x,y)}{\diff x\diff y}\,.
\label{eq:dx}
\end{equation}
This immediately translates into
\begin{equation}
\delta(x)
=\left(\overline{|\mathcal{M}_G^\text{LO}(x)|^2}\right)^{-1}
\int_{-\beta}^{\beta}\diff y\,
\delta(x,y)
\overline{|\mathcal{M}_G^\text{LO}(x,y)|^2}\,,
\label{eq:dxG}
\end{equation}
where we defined
\begin{equation}
\begin{split}
&\overline{|\mathcal{M}_G^\text{LO}(x)|^2}
\equiv\int_{-\beta}^{\beta}\diff y\,\overline{|\mathcal{M}_G^\text{LO}(x,y)|^2}\\
&=\frac{4\beta}3\left(1+\frac{\nu^2}{2x}\right)\left\{|G_\text{E}(\Delta_M^2x)|^2+2\rho x|G_\text{M}(\Delta_M^2x)|^2\right\}.
\end{split}
\label{eq:MGLOx}
\end{equation}
After neglecting the electric form factor relative to the magnetic one, we obtain
\begin{equation}
\delta(x)\simeq
\left[\frac{8\beta}3\left(1+\frac{\nu^2}{2x}\right)\right]^{-1}\int_{-\beta}^{\beta}\diff y\,\delta(x,y)\left(1+y^2+\frac{\nu^2}{x}\right).
\label{eq:dxapprox}
\end{equation}
In the following sections we discuss the individual contributions.

\section{Virtual radiative corrections}
\label{sec:virt}

\begin{figure}[t]
\vspace{-2mm}
\centering
\subfloat[][]{
\includegraphics[width=0.45\columnwidth]{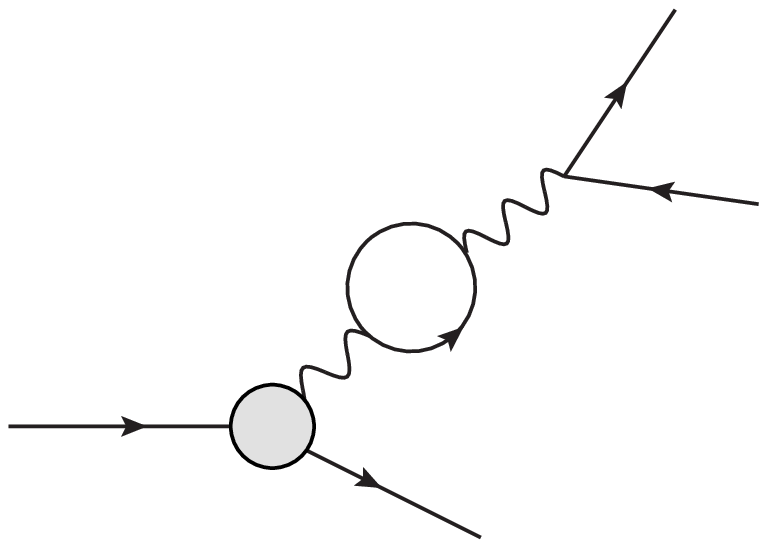}
\label{fig:virta}
}
\subfloat[][]{
\includegraphics[width=0.45\columnwidth]{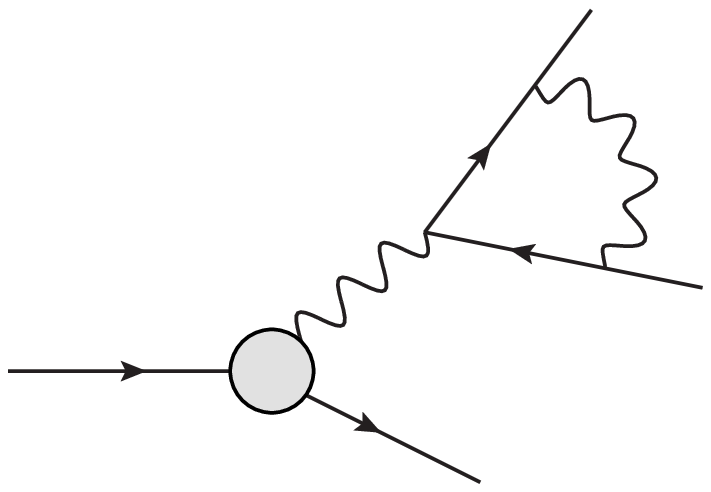}
\label{fig:virtb}
}

\subfloat[][]{
\includegraphics[width=0.45\columnwidth]{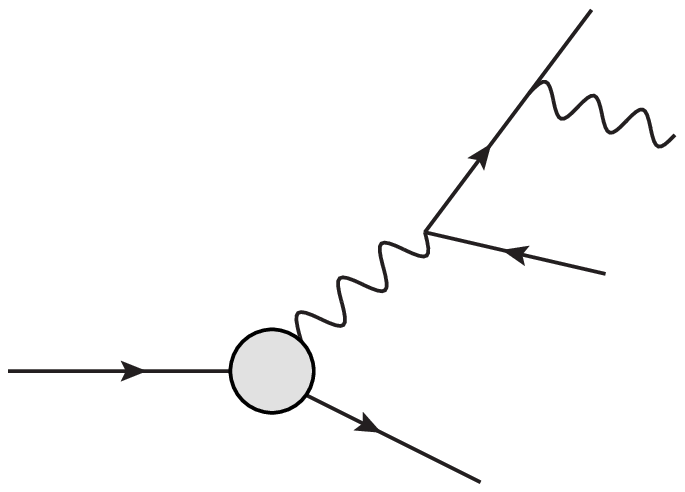}
\label{fig:BS}
}
\subfloat[][]{
\includegraphics[width=0.45\columnwidth]{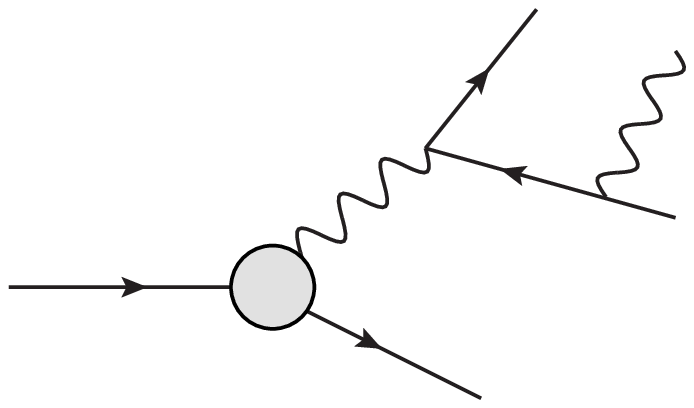}
\label{fig:BS2}
}

\subfloat[][]{
\includegraphics[width=0.45\columnwidth]{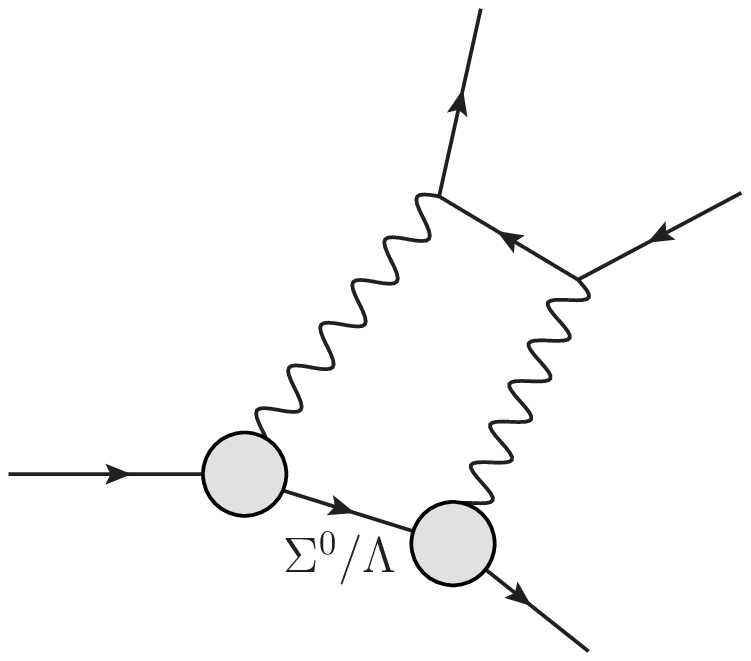}
\label{fig:1gIR}
}
\subfloat[][]{
\includegraphics[width=0.45\columnwidth]{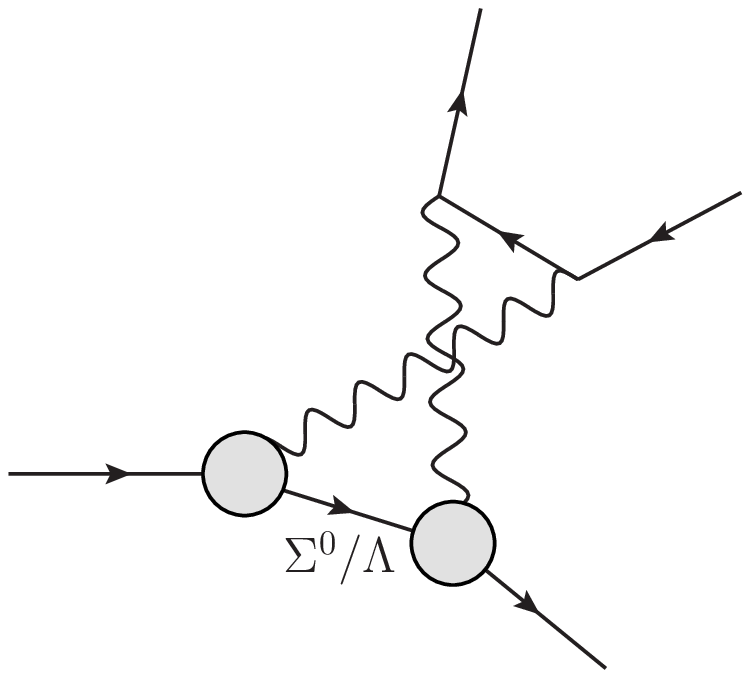}
\label{fig:1gIR2}
}

\subfloat[][]{
\includegraphics[width=0.45\columnwidth]{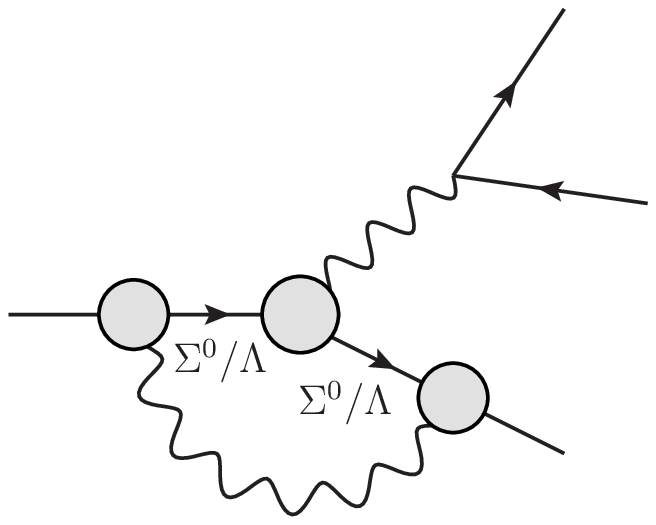}
\label{fig:virt_bar}
}
\vspace{4mm}
\caption{
\label{fig:diagrams}
NLO QED radiative corrections for the decay $\Sigma^0\to\Lambda e^+e^-$: a) lepton-loop vacuum-polarization insertion, b) correction to the QED vertex, c) \& d) bremsstrahlung, e) \& f) one-loop one-photon-irreducible (1$\gamma$IR) contributions, g) $\Sigma^0\Lambda\gamma$ vertex correction.
In the 1$\gamma$IR contribution each diagram comes in two variants: with $\Sigma^0$ or $\Lambda$ exchanged.
Similarly, there are four diagrams contributing to the transition-form-factor correction g).
}
\end{figure}

We get the virtual radiative corrections, $\delta^\text{virt}$, from the interference terms of the LO diagram shown in Fig.~\ref{fig:LO} and the NLO one-loop diagrams of Figs.~\ref{fig:virta} and~\ref{fig:virtb}.
We recall that the LO expression is given in Eq.~(\ref{eq:MLO}) and the general prescription is provided in Eq.~(\ref{eq:M}).
The result for this interference can be written as
\begin{equation}
\begin{split}
&\delta^\text{virt}(x,y)
=2\operatorname{Re}\biggl\{-\tilde\Pi(\Delta_M^2x)+F_1(\Delta_M^2x)\biggr.\\
&+\left.F_2(\Delta_M^2x)\,\frac{|G_\text{E}(\Delta_M^2x)|^2+2\rho x|G_\text{M}(\Delta_M^2x)|^2}{\overline{|\mathcal{M}_G^\text{LO}(x,y)|^2}}\right\}.
\end{split}
\label{eq:dvirt}
\end{equation}
If we take into account again that the electric form factor is suppressed, we arrive at the expression
\begin{equation}
\delta^\text{virt}(x,y)
\simeq
2\operatorname{Re}\left\{-\tilde\Pi(\Delta_M^2x)+F_1(\Delta_M^2x)+\frac{2F_2(\Delta_M^2x)}{1+y^2+\frac{\nu^2}{x}}\right\},
\label{eq:dvirtapprox}
\end{equation}
which is exactly the form one finds in the case of the Dalitz decay of the neutral pion~\cite{Mikaelian:1972yg,Husek:2015sma} and which is numerically more than a satisfactory approximation.
This finding is further supported by the fact that compared to the size of the effects stemming from $\tilde\Pi(q^2)$ and $F_1(q^2)$, the quantity $F_2(q^2)$ is numerically negligible in the whole kinematically allowed region of $q^2$.
Moreover, if we recall the form of $\overline{|\mathcal{M}_G^\text{LO}(x)|^2}$ from Eq.~(\ref{eq:MGLOx}), it is straightforward to see that the contribution to the correction to the one-fold differential decay width takes --- of course, in both cases of Eqs.\ (\ref{eq:dvirt}) and (\ref{eq:dvirtapprox}) --- the form independent of the electric and magnetic form factors:
\begin{equation}
\begin{split}
\delta^\text{virt}(x)
&=2\operatorname{Re}\biggl\{
-\tilde\Pi(\Delta_M^2x)\\
&+\left.F_1(\Delta_M^2x)+\frac32\left(1+\frac{\nu^2}{2x}\right)^{-1}F_2(\Delta_M^2x)\right\}.
\end{split}
\label{eq:dvirtx}
\end{equation}
We should now evaluate these terms at NLO.

Considering the correction stemming from the diagram in Fig.~\ref{fig:virta}, we find for the (on-shell renormalized) vacuum-polarization insertion induced by a single leptonic loop (flavor $\ell$)
\begin{equation}
\overline\Pi_\ell(\Delta_M^2x)
=\frac{\alpha}{\pi}
\left\{\frac89-\frac{\beta_\ell^2}3+\frac{\beta_\ell}2\left(1-\frac{\beta_\ell^2}3\right)\log[-\gamma_\ell+i\epsilon]\right\},
\label{eq:VPIl}
\end{equation}
with $\beta_\ell\equiv\beta_\ell(x)\equiv\sqrt{1-\nu_\ell^2/x}$, $\nu_\ell\equiv2m_\ell/\Delta_M$ and
\begin{equation}
\gamma_\ell\equiv\gamma_\ell(x)\equiv\frac{1-\beta_\ell(x)}{1+\beta_\ell(x)}\,.
\end{equation}
In what follows, it is enough to take into account only the electron- and muon-loop contributions,
\begin{equation}
\overline\Pi(q^2)
=\sum_{\ell\in\{\text{e},\mu\}}\overline\Pi_\ell(q^2)\,,
\end{equation}
the latter mentioned having only a cosmetic effect on the presented numerical results.
This is due to the fact that the exchanged invariant mass $\sqrt{q^2}$ does not exceed $\Delta_M\approx77$\,MeV and is thus way below the two-muon threshold.
Let us remark that independently of the considered processes, the contribution with the lightest fermion is numerically of the largest importance.
After summing the whole geometric series of one-loop insertions ($-\tilde\Pi(q^2)=1/[1+\overline\Pi(q^2)]$), squaring the amplitude and subtracting the LO part, we get for the correction connected to the vacuum polarization
\begin{equation}
\delta_\Pi^\text{virt}(x,y)
=\delta_\Pi^\text{virt}(x)
=\frac1{|1+\overline\Pi(\Delta_M^2x)|^2}-1\,.
\end{equation}
This term can then be used (and we do so) instead of the one-loop interference term $-2\operatorname{Re}\tilde\Pi(\Delta_M^2x)$ in Eq.~(\ref{eq:dvirt}) and further on.

For the electromagnetic form factors $F_1(q^2)$ and $F_2(q^2)$ stemming from the QED vertex correction in Fig.~\ref{fig:virtb} we have at NLO (on-shell renormalized)
\begin{equation}
\begin{split}
&\overline{F_1^\text{NLO}}(\Delta_M^2x)
=\frac{\alpha}{\pi}
\biggl\{
-1-\frac{1+2\beta^2}{4\beta}\log(-\gamma+i\epsilon)-\frac{1+\beta^2}{2\beta}\biggr.\\
&\times\bigg[\text{Li}_2(1-\gamma)+\frac14\log^2(-\gamma+i\epsilon)
-\frac{\pi^2}{4}-i\pi\log(1-\gamma)\bigg]\\
&+\biggl.\left[1+\frac{1+\beta^2}{2\beta}\log(-\gamma+i\epsilon)\right]\log\frac m{\Lambda}
\biggr\}
\end{split}
\label{eq:F1}
\end{equation}
and
\begin{equation}
F_2^\text{NLO}(\Delta_M^2x)
=\frac{\alpha}{\pi}\frac{\nu^2}{4x\beta}\log{(-\gamma+i\epsilon)}\,.
\label{eq:F2}
\end{equation}
In the above formulae, $\text{Li}_2$ stands for the dilogarithm and $\Lambda$ is the infrared cut-off.
Note that we quote here the full expression valid in all kinematic regimes.
In order to extract the real parts from Eqs.~(\ref{eq:F1}) and (\ref{eq:F2}), in the kinematically allowed region where $\Delta_M^2x\ge4m^2$ we use $\log(-\gamma+i\epsilon)=\log(\gamma)+i\pi$, since $0\le\gamma\le1$.
It is then straightforward to see that the real part of $\overline{F_1^\text{NLO}}$ indeed includes the Coulomb term proportional to $-\pi^2/2$.

For completeness, our final expressions for the virtual radiative corrections take the form
\begin{equation}
\begin{split}
&\delta^\text{virt}(x,y)
=\frac1{|1+\overline\Pi(\Delta_M^2x)|^2}-1
+2\operatorname{Re}\biggl\{\overline{F_1^\text{NLO}}(\Delta_M^2x)\biggr.\\
&+\left.F_2^\text{NLO}(\Delta_M^2x)\,\frac{|G_\text{E}(\Delta_M^2x)|^2+2\rho x|G_\text{M}(\Delta_M^2x)|^2}{\overline{|\mathcal{M}_G^\text{LO}(x,y)|^2}}\right\},
\end{split}
\label{eq:dvirtfinal}
\end{equation}
and, using Eq.~(\ref{eq:dxapprox}) or based on Eq.~(\ref{eq:dvirtx}),
\begin{equation}
\begin{split}
&\delta^\text{virt}(x)
=\frac1{|1+\overline\Pi(\Delta_M^2x)|^2}-1\\
&+2\operatorname{Re}\left\{
\overline{F_1^\text{NLO}}(\Delta_M^2x)+\frac32\left(1+\frac{\nu^2}{2x}\right)^{-1}F_2^\text{NLO}(\Delta_M^2x)\right\}.
\end{split}
\label{eq:dvirtxfinal}
\end{equation}

\section{Bremsstrahlung}
\label{sec:BS}

Concerning the notation, we stick to the one provided previously in Refs.~\cite{Mikaelian:1972yg,Husek:2015sma,Husek:2017vmo}.
The two diagrams which contribute to the bremsstrahlung of the process $\Sigma^0\to\Lambda e^+e^-$ are shown in Figs.~\ref{fig:BS} and \ref{fig:BS2}.
Besides other effects, their presence is essential to cancel the IR divergence stemming from the virtual corrections depicted in Fig.~\ref{fig:virtb}.
The corresponding invariant matrix element can be written in the form
\begin{equation}
\begin{split}
&i\mathcal{M}_\text{BS}
=-\frac{i^5e^3}{(k+q_1+q_2)^2+i\epsilon}\\
&\times[\bar u_\Lambda(\vec p_2) G_\mu(p_1-p_2) u_\Sigma(\vec p_1)]
[\bar u(\vec q_1)I^{\mu\rho}v(\vec q_2)]
\,\epsilon_\rho^*(k)\,,
\end{split}
\label{eq:MBS}
\end{equation}
where
\begin{equation}
\begin{split}
I^{\alpha\beta}
=\gamma^\beta\frac{(\slashed k+\slashed q_1+m)}{2\,k\cdot q_1+i\epsilon}\gamma^\alpha
-\gamma^\alpha\frac{(\slashed k+\slashed q_2-m)}{2\,k\cdot q_2+i\epsilon}\gamma^\beta\,.
\end{split}
\end{equation}
Here, we use $k$ for the four-momentum of the bremsstrahlung photon.
Inasmuch as an additional particle comes into play, it is convenient to introduce a new kinematic variable which stands for the invariant mass squared of the pair formed by the photon and the $\Lambda$ hyperon:
\begin{equation}
s_\gamma
\equiv(k+p_2)^2\,.
\end{equation}
It is a counterpart to $s\equiv(q_1+q_2)^2=\Delta_M^2x$.

The form factors $G_1$ and $G_2$ are translated into $G_\text{M}$ and $G_\text{E}$ via Eq.~\eqref{eq:G12}.
They are further approximated using the linear expansion shown in Eq.~\eqref{eq:expandGEGM}.
This yields
\begin{align}
G_\text{M}((k+q_1+q_2)^2)
&\simeq G_\text{M}(s)\bigg\{1+\frac16\langle r_\text{M}^2\rangle[2k\cdot(q_1+q_2)]\bigg\}\,,
\label{eq:GMBS}\\
G_\text{E}((k+q_1+q_2)^2)
&\simeq G_\text{E}(s)\bigg\{1+\frac{2k\cdot(q_1+q_2)}{s}\bigg\}\,.
\label{eq:GEBS}
\end{align}
In what follows we consider the above form of the form factors to be used in the evaluation of the bremsstrahlung correction.

The contribution of the bremsstrahlung to the NLO two-fold differential decay width can be written as
\begin{equation}
\frac{\diff^2\Gamma_\text{BS}^\text{NLO}(x,y)}{\diff x\diff y}
=\frac{1}{2M_\Sigma}\frac{\pi^3\Delta_M^2\lambda^{\frac12}(x)}{16(2\pi)^8M_\Sigma^2}
\int J\Big\{\overline{|\mathcal{M}_\text{BS}|^2}\Big\}\diff s_\gamma\,.
\label{eq:dBS}
\end{equation}
The above used operator $J$ is defined for an arbitrary invariant $f(k,p_2)$ of the momenta $k$ and $p_2$ as follows:
\begin{equation}
\begin{split}
&J\{f(k,p_2)\}\\
&=\frac 1{2\pi}\int\frac{\diff^3k}{k_0}\frac{\diff^3p_2}{p_{2,0}}f(k,p_2)\,\delta^{(4)}(p_1-q_1-q_2-p_2-k)\,.
\end{split}
\end{equation}
Being on the mass shell ($k^2=0,\,p_2^2=M_\Lambda^2$) and in the reference system where $\vec p_1-\vec q_1-\vec q_2=0(=\vec p_2+\vec{k}\equiv\vec{r})$, we find
\begin{equation}
J[f(k,p_2)]
\stackrel{(\vec{r}=0)}{=}
\frac 1{4\pi}\frac{\tilde\omega}\omega\int\diff\Omega_{\vec{k}}\,f(k,\tilde{k}(M_\Lambda^2))\big|_{|\vec{k}|=\frac{{\tilde\omega}}2}\,,
\end{equation}
where $\omega\equiv\sqrt{s_\gamma}$ and ${\tilde\omega}\equiv(s_\gamma-M_\Lambda^2)/\sqrt{s_\gamma}$\,.
We used $\tilde{k}(M_\Lambda^2)$ to mark the four-momentum of the particle with the mass $M_\Lambda$ and with the momentum $-\vec{k}$, i.e.\ when $k=(k_0,\vec{k})$, then for the four-vector $\tilde k(M_\Lambda^2)$ we write
\begin{equation}
\tilde k(M_\Lambda^2)=\left(\sqrt{|\vec{k}|^2+M_\Lambda^2},-\vec{k}\right)
=p_2\big|_{\vec p_2=-\vec{k}}\,.
\end{equation}
We can come back to the invariant form through
\begin{equation}
k_0+p_{2,0}
\stackrel{(\vec{r}=0)}{=}\sqrt{(k+p_2)^2}=\sqrt{s_\gamma}=\omega
\end{equation}
or for example due to
\begin{equation}
q_{1,0}=\frac{(k_0+p_{2,0})\,q_{1,0}}{k_0+p_{2,0}}
\stackrel{(\vec{r}=0)}{=}\frac{(k+p_2)\cdot q_1}{\omega}\,.
\end{equation}
Together with $x$ and $y$ (kinematic variables at LO with the same meaning also at NLO) and $s_\gamma$, two more independent kinematic variables are necessary.
We define
\begin{equation}
A\equiv k\cdot q_2\,,\quad
B\equiv k\cdot q_1\,,\quad
E\equiv (k+q_1+q_2)^2\,,
\end{equation}
where e.g.\ $E$ can be expressed in terms of $A$ and $B$ as $E=\Delta_M^2x+2A+2B$.
Finally, the bremsstrahlung correction reads
\begin{equation}
\begin{split}
&\delta^\text{BS}(x,y)\\
&=\frac\alpha\pi
\frac14
\left[\frac{(1-x)}{\rho x^2}\,\overline{|\mathcal{M}_G^\text{LO}(x,y)|^2}\right]^{-1}
\int\frac1{2e^6}J\Big\{\overline{|\mathcal{M}_\text{BS}|^2}\Big\}\diff s_\gamma\,.
\end{split}
\label{eq:deltaBS}
\end{equation}

The full result for the matrix element squared of the bremsstrahlung correction $\overline{|\mathcal{M}_\text{BS}|^2}$ is rather lengthy and it makes not much sense to present it here.
Nevertheless, all the terms necessary to numerically evaluate $\delta^\text{BS}(x,y)$ are presented in \ref{app:J} and numerical results given later in Table~\ref{tab:deltaSym} of Section~\ref{sec:res} correspond to using the expansions~(\ref{eq:GMBS}) and (\ref{eq:GEBS}).
In what follows we present a very simple --- however numerically satisfactory --- form of $\overline{|\mathcal{M}_\text{BS}|^2}$.
Assuming that $G_\text{E}(s)\simeq0$, $G_\text{M}((k+q_1+q_2)^2)\simeq G_\text{M}(s)$, $\rho\ll1$ and neglecting sub-leading terms in $\nu^2$, the (IR-)convergent part (to be integrated {\em numerically} over $s_\gamma$) of the bremsstrahlung matrix element squared can be written as follows:
\begin{equation}
\begin{split}
&\frac1{2e^6}\overline{|\mathcal{M}_\text{BS}|^2}\Big|_\text{C}\\
&\simeq
2|G_\text{M}(s)|^2\bigg\{
\frac1E
+\frac1{E-4\hat M^2}
-2\Delta_M^2\frac1{E^2}
-\frac1A
\\&
+\frac{(k\cdot p_2)^2}{4\hat M^2}\frac1{AB}
-\Delta_M^2s\frac1{AE^2}
+\frac12[s+2\sqrt{\rho}(\Delta_m^2-2k\cdot p_2)]\frac1{AE}
\\&
-\frac1{4\hat M^2}[8\hat M^4+4\hat M^2\Delta_m^2+\Delta_m^4+4k\cdot p_2(2\hat M^2+k\cdot p_2)]
\\&
\times\frac1{A(E-4\hat M^2)}
-\frac{\nu^2\rho}2\frac{k\cdot p_2}{s}(\Delta_m^2+k\cdot p_2)\frac1{A^2}
\\&
-\frac{\nu^2}{8}[\Delta_M^4+\rho(s-\Delta_m^2)^2-4\rho\,k\cdot p_2(s-\Delta_m^2-k\cdot p_2)]\frac1{A^2E}
\\&
+\frac{\nu^2\rho}8(4\hat M^2+\Delta_m^2+2k\cdot p_2)^2\frac1{A^2(E-4\hat M^2)}
\bigg\}
+(q_1\leftrightarrow q_2)
\,.
\end{split}
\label{eq:MBSconv}
\end{equation}
Above, we used $\Delta_m^2\equiv-2p_1\cdot(q_1-q_2)$.
Note that the dimensionless variables $x$ and $y$ are related to $s$ and $\Delta_m^2$ in the following manner: $s=\Delta_M^2x$ and $\Delta_m^2\equiv y\lambda^{\frac12}(x)$; see Eq.~(\ref{eq:lambdax}) for the definition of $\lambda(x)$.
The second half of the expression --- denoted as $(q_1\leftrightarrow q_2)$ --- is related to the sign change of $y$ (or $\Delta_m^2$) and to the change $(A\leftrightarrow B$) performed on the first half of Eq.~\eqref{eq:MBSconv}.
For instance, it can be obtained when the first part is inserted into the operator $J$, the expressions from \ref{app:J} are substituted and afterwards one uses $(\Delta_m^2\leftrightarrow-\Delta_m^2)$.
The simplified result \eqref{eq:MBSconv} does not require the knowledge of all the basic integrals listed in \ref{app:J}:
Indeed, some of them are only necessary when the exact $\overline{|\mathcal{M}_\text{BS}|^2}\big|_\text{C}$ at NLO is evaluated as was done to generate Table~\ref{tab:deltaSym}.
The full form of the (IR-)divergent part (to be integrated {\em analytically}) then reads:
\begin{equation}
\begin{split}
&\frac1{2e^6}\overline{|\mathcal{M}_\text{BS}|^2}\Big|_\text{D}
=4\frac{(1-x)}{\rho x^2}\,\overline{|\mathcal{M}_G^\text{LO}(x,y)|^2}\\
&\times\left[\left(1-\frac{\nu^2}{2x}\right)\frac s4\frac1{AB}
-\frac{m^2}{4}\left(\frac1{A^2}+\frac1{B^2}\right)\right].
\end{split}
\label{eq:MBSdiv}
\end{equation}
After substituting Eqs.\ (\ref{eq:JA2div}) and (\ref{eq:JABdiv}) into Eq.~(\ref{eq:MBSdiv}) and further into Eq.~(\ref{eq:deltaBS}), the terms proportional to $\ln(m/\Lambda)$ cancel with those in Eq.~(\ref{eq:F1}), i.e.\ $2\operatorname{Re}\overline{F_1^\text{NLO}}(\Delta_M^2x)+\delta^\text{BS}(x,y)\big|_\text{D}$ is IR-finite, using $(1-\frac{\nu^2}{2x})=\beta\frac{1+\beta^2}{2\beta}$\,.
In order to get an approximate result of the bremsstrahlung correction, one should simply substitute for $\frac1{2e^6}\overline{|\mathcal{M}_\text{BS}|^2}$ the sum of terms (\ref{eq:MBSconv}) and (\ref{eq:MBSdiv}).

\section{One-photon-irreducible virtual radiative corrections}
\label{sec:1gIR}

The diagrams of Figs.~\ref{fig:LO} and \ref{fig:virta}-\ref{fig:BS2} contain only the hadronic form factors for the transition from the Sigma to the Lambda hyperon.
In contrast to this, the remaining diagrams of Figs.~\ref{fig:1gIR}, \ref{fig:1gIR2} and \ref{fig:virt_bar} contain also other hadronic form factors.
Strictly speaking, what is denoted by ``$\Sigma^0$/$\Lambda$'' in Fig.~\ref{fig:diagrams} could even be other intermediate baryonic states with strangeness.
Yet, we will show in the following that the 1$\gamma$IR contributions are negligible.
For this purpose, it is not required to account in a detailed and utterly correct way for all possible hadronic effects.
A sample calculation should be sufficient.
Therefore, we restrict ourselves indeed to the diagrams shown in Figs.\ \ref{fig:1gIR}, \ref{fig:1gIR2} and \ref{fig:virt_bar} and use form factors that are sufficiently realistic.
To strengthen our statement we will explore a variety of form factors.

For the 1$\gamma$IR contribution (Figs.~\ref{fig:1gIR} and \ref{fig:1gIR2}), we need to calculate four box diagrams.
The matrix element can be separated into a baryonic ($B_{\mu\nu}$) and leptonic ($L_{\alpha\beta}$) part
\begin{equation}
i\mathcal{M}^{1\gamma\text{IR}}
=i^8e^4\hskip-1mm\int\hskip-1mm\frac{\diff^4l}{(2\pi)^4}\,B_{\mu\nu}\frac{g^{\mu\alpha}}{(l-p_2)^2}\frac{g^{\nu\beta}}{(l-p_1)^2}L_{\alpha\beta}\,,
\end{equation}
where the photon propagators are shown explicitly (for brevity we drop the `$+i\epsilon$' parts).

The leptonic part can be written as follows:
\begin{equation}
\begin{split}
L_{\alpha\beta}
&=\bar u_e(\vec q_1)\gamma_\alpha\frac{1}{-\slashed l+\slashed p_2+\slashed q_1-m}\gamma_\beta v_e(\vec q_2)\\
&+\bar u_e(\vec q_1)\gamma_\beta\frac{1}{\slashed l-\slashed p_2-\slashed q_2-m}\gamma_\alpha v_e(\vec q_2)\,.
\end{split}
\label{eq:L}
\end{equation}
The cross term is substantial to attain gauge invariance, which manifests itself as
\begin{equation}
(l-p_2)^\alpha L_{\alpha\beta}=0=(p_1-l)^\beta L_{\alpha\beta}\,.
\label{eq:L_gauge}
\end{equation}
This can be seen when we artificially rewrite Eq.~(\ref{eq:L}) with the Dirac equation at hand,
\begin{equation}
\begin{split}
L_{\alpha\beta}
&=-\bar u_e(\vec q_1)(\gamma_\alpha-\slashed q_1+m)\frac{1}{\slashed l-\slashed p_2-\slashed q_1+m}\gamma_\beta v_e(\vec q_2)\\
&+\bar u_e(\vec q_1)\gamma_\beta\frac{1}{\slashed l-\slashed p_2-\slashed q_2-m}(\gamma_\alpha-\slashed q_2-m) v_e(\vec q_2)\,,
\end{split}
\end{equation}
the form of which is suited to show the first equality in Eq.~(\ref{eq:L_gauge}), or, using the energy-momentum-conservation relation $p_1=p_2+q_1+q_2$,
\begin{equation}
\begin{split}
L_{\alpha\beta}
&=\bar u_e(\vec q_1)\gamma_\alpha\frac{1}{-\slashed l+\slashed p_1-\slashed q_2-m}(\gamma_\beta-\slashed q_2-m) v_e(\vec q_2)\\
&-\bar u_e(\vec q_1)(\gamma_\beta-\slashed q_1+m)\frac{1}{-\slashed l+\slashed p_1-\slashed q_1+m}\gamma_\alpha v_e(\vec q_2)
\end{split}
\end{equation}
to obtain the second equality.

Regarding the baryonic part, for the purpose of treating the one-loop diagrams we should consider to generalize some of the previous definitions.
Instead of Eq.~(\ref{eq:Gon}) we will now use
\begin{equation}
\langle Y|j_\mu|X\rangle
=e\bar u_Y(\vec p_2) G_\mu^{XY}(p_1-p_2) u_X(\vec p_1)\,,
\end{equation}
where $X,Y\in\{\Lambda,\Sigma^0\}$, and
\begin{equation}
G_\mu^{XY}(q)
\equiv
\bigg[\gamma_\mu-\frac{q_\mu}{q^2}\slashed q\bigg]G_1^{XY}({q^2})
-\frac{i\sigma_{\mu\nu}q^\nu}{M_X+M_Y}G_2^{XY}(q^2)\,.
\label{eq:Goff}
\end{equation}
We can see that this definition of baryonic electromagnetic form factors is manifestly gauge invariant, which now holds also off-shell.
On-shell it reduces to Eq.~(\ref{eq:Gon}).
In other words
\begin{equation}
q^\mu G_\mu^{XY}(q)=0\,.
\end{equation}
The baryonic part can then be understood as a sum of two contributions
\begin{equation}
B_{\mu\nu}
=B_{\mu\nu}^{\Sigma^0\Lambda,\Lambda\Lambda}
+B_{\mu\nu}^{\Sigma^0\Sigma^0,\Sigma^0\Lambda}\,,
\label{eq:Bcontr}
\end{equation}
where we introduced
\begin{equation}
B_{\mu\nu}^{\Sigma^0X,X\Lambda}
\equiv\bar u_\Lambda(\vec p_2)G_\mu^{X\Lambda}(l-p_2)\frac{1}{\slashed l-M_X}G_\nu^{\Sigma^0X}(p_1-l)u_\Sigma(\vec p_1)\,.
\label{eq:B_ABBC}
\end{equation}
Looking at previous equations one can easily check that the baryonic part is gauge invariant
\begin{equation}
(l-p_2)^\mu B_{\mu\nu}=0=(p_1-l)^\nu B_{\mu\nu}\,,
\label{eq:B_gauge}
\end{equation}
which already holds for the separate contributions in Eq.~(\ref{eq:Bcontr}).

Having the conservation of the electromagnetic current (\ref{eq:L_gauge}) in mind, we can somewhat simplify the baryonic part $B_{\mu\nu}$ of the matrix element.
Exploiting the operator identity $2AB=[A,B]+\{A,B\}$ we can write
\begin{equation}
[\gamma^\nu,\slashed p_1-\slashed l]u_\Sigma(\vec p_1)L_{\mu\nu}
=2\gamma^\nu(M_\Sigma-\slashed l)u_\Sigma(\vec p_1)L_{\mu\nu}
\label{eq:commL1}
\end{equation}
and
\begin{equation}
\bar u_\Lambda(\vec p_2)[\gamma^\mu,\slashed l-\slashed p_2]L_{\mu\nu}
=2\bar u_\Lambda(\vec p_2)(M_\Lambda-\slashed l)\gamma^\mu L_{\mu\nu}\,.
\label{eq:commL2}
\end{equation}
Above, we used the Dirac equation and the fact that the anticommutator part $\{A,B\}$ disappears due to Eq.~(\ref{eq:L_gauge}).
Effectively, we can thus take only
\begin{equation}
G_\mu^{XY}(q)
\equiv G_1^{XY}({q^2})\gamma_\mu
+\frac{[\gamma_\mu,\slashed q]}{2(M_X+M_Y)}G_2^{XY}(q^2)
\end{equation}
instead of the full form (\ref{eq:Goff}) inside of Eq.~(\ref{eq:B_ABBC}), or even more explicitly and using Eqs.~(\ref{eq:commL1}-\ref{eq:commL2}), we write
\begin{equation}
\begin{split}
&\tilde{B}_{\mu\nu}^{\Sigma^0X,X\Lambda}\\
&=\bar u_\Lambda(\vec p_2)
\bigg\{
G_1^{X\Lambda}\big((l-p_2)^2\big)
-\frac{(\slashed l-M_\Lambda)}{M_X+M_\Lambda}G_2^{X\Lambda}\big((l-p_2)^2\big)
\bigg\}\\
&\times\gamma_\mu
\frac{\slashed l+M_X}{l^2-M_X^2}
\gamma_\nu\\
&\times\bigg\{
G_1^{\Sigma^0X}\big((l-p_1)^2\big)
-\frac{(\slashed l-M_\Sigma)}{M_X+M_\Sigma}G_2^{\Sigma^0X}\big((l-p_1)^2\big)
\bigg\}
u_\Sigma(\vec p_1)\,.
\end{split}
\label{eq:Btilde}
\end{equation}
We see, that the off-shell redefinition~(\ref{eq:Goff}) of Eq.~(\ref{eq:Gon}) does not affect, at the end of the day, the result (\ref{eq:Btilde}) due to the conservation laws specified above.

The LO amplitude of the $\Sigma^0\to\Lambda e^+e^-$ decay (\ref{eq:MLO}) can be written in the following way:
\begin{equation}
\begin{split}
i\mathcal{M}^\text{LO}
&=\frac{i^3e^2}{\Delta_M^2x}\big[\bar u_\Lambda(\vec p_2)\gamma_\sigma u_\Sigma(\vec p_1)\big]\big[\bar u_e(\vec q_1)\gamma_\tau v_e(\vec q_2)\big]\\
&\times\bigg\{G_\text{M}(\Delta_M^2x)\,g^{\sigma\tau}-{G_2(\Delta_M^2x)}\frac{p_1^\sigma p_2^\tau}{M_\Sigma\hat M}\bigg\}
\,.
\end{split}
\end{equation}
Its interference with the 1$\gamma$IR amplitude represented by the box diagrams can be obtained by separately treating the leptonic and baryonic parts.
For the leptonic part we can write (summing over the final-state degrees of freedom)
\begin{equation}
\sum_\text{spins}L_{\alpha\beta}\big[\bar u_e(\vec q_1)\gamma_\tau v_e(\vec q_2)\big]^*
=\tilde{L}_{\alpha\beta\tau}(q_1,q_2)-\tilde{L}_{\alpha\beta\tau}(q_2,q_1)\,,
\label{eq:Lab_LL}
\end{equation}
where
\begin{equation}
\begin{split}
&\tilde{L}_{\alpha\beta\tau}(q_1,q_2)
\equiv\sum_\text{spins}\tilde{L}_{\alpha\beta}(q_1,q_2)\big[\bar u_e(\vec q_1)\gamma_\tau v_e(\vec q_2)\big]^*\\
&=\frac{1}{(l-p_2-q_1)^2-m^2}
\bigg\{
2q_{1\alpha}\text{Tr}\big\{(\slashed q_1+m)\gamma_\beta(\slashed q_2-m)\gamma_\tau\big\}\\
&-(l-p_2)^\rho
\text{Tr}\big\{(\slashed q_1+m)\gamma_\alpha\gamma_\rho\gamma_\beta(\slashed q_2-m)\gamma_\tau\big\}
\bigg\}\,.
\end{split}
\end{equation}
Above we used $\tilde{L}_{\alpha\beta}(q_1,q_2)$ for the first term of $L_{\alpha\beta}$ as seen in Eq.~(\ref{eq:L}).
The validity of Eq.~(\ref{eq:Lab_LL}) can be technically checked exploiting
\begin{equation}
\begin{split}
\text{Tr}\big\{\gamma_\alpha\gamma_\beta\dots\gamma_\rho\gamma_\sigma\big\}
&=\text{Tr}\big\{\gamma_\sigma\gamma_\rho\dots\gamma_\beta\gamma_\alpha\big\}\,.
\end{split}
\end{equation}
For the treatment of the loop integral, it will be convenient to explicitly extract the loop momentum in the following way:
\begin{equation}
\tilde{L}_{\alpha\beta\tau}(q_1,q_2)
=\frac{\text{Tr}_{\alpha\beta\tau}^{\text{L}}(q_1,q_2)}{(l-p_2-q_1)^2-m^2}
-\frac{l^\kappa\text{Tr}_{\alpha\kappa\beta\tau}^{\text{L}}(q_1,q_2)}{(l-p_2-q_1)^2-m^2}\,.
\label{eq:Labt}
\end{equation}
The traces in the numerators are then simply defined as
\begin{align}
\begin{split}
\text{Tr}_{\alpha\beta\tau}^{\text{L}}(q_1,q_2)
&\equiv\text{Tr}\big\{(\slashed q_1+m)(2q_{1\alpha}+\gamma_\alpha\slashed p_2)\gamma_\beta(\slashed q_2-m)\gamma_\tau\big\}\,,
\end{split}\\
\text{Tr}_{\alpha\kappa\beta\tau}^{\text{L}}(q_1,q_2)
&\equiv\text{Tr}\big\{(\slashed q_1+m)\gamma_\alpha\gamma_\kappa\gamma_\beta(\slashed q_2-m)\gamma_\tau\big\}\,.
\end{align}
Similarly, one of the two contributions (in the sense of Eq.~(\ref{eq:Bcontr}) and defined in Eq.~(\ref{eq:Btilde})) to the baryonic part can be written in the following form:
\begin{equation}
\begin{split}
&\tilde{B}_{\mu\nu\sigma}^{X}
=\sum_\text{spins}\tilde{B}_{\mu\nu}^{\Sigma^0X,X\Lambda}\big[\bar u_\Lambda(\vec p_2)\gamma_\sigma u_\Sigma(\vec p_1)\big]^*\\
&=\frac1{l^2-M_X^2}\sum_{i=1}^8 \beta_i^X \text{Tr}\big\{(\slashed p_2+M_\Lambda)T_{\mu\nu}^i(\slashed p_1+M_\Sigma)\gamma_\sigma\big\}\,,
\end{split}
\label{eq:BX}
\end{equation}
with the coefficients $\beta_i^X$ and matrices $T_{\mu\nu}^i$ listed in \ref{app:betaT}.
It is also convenient to define the trace
\begin{equation}
\text{Tr}_{\mu..\nu,\sigma}^{\text{B}}
\equiv\text{Tr}\big\{(\slashed p_2+M_\Lambda)\gamma_{\mu..\nu}(\slashed p_1+M_\Sigma)\gamma_\sigma\big\}\,,
\end{equation}
where we used the short-hand notation for a product of $\gamma$-matrices:
\begin{equation}
\gamma_{\rho\sigma\dots\tau}
\equiv\gamma_\rho\gamma_\sigma\dots\gamma_\tau\,.
\label{eq:gamma_prod}
\end{equation}

A contribution of the box diagrams to the NLO matrix element squared of the $\Sigma^0\to\Lambda e^+e^-$ decay can then be expressed as the interference
\begin{equation}
\begin{split}
&\overline{|\mathcal{M}^\text{LO+NLO}|^2}\Big|_{1\gamma\text{IR}}
\equiv2\operatorname{Re}\sum_\text{spins}\frac12\,\mathcal{M}^{1\gamma\text{IR}}\mathcal{M}^\text{LO*}\\
&=\operatorname{Re}\bigg\{
\frac{ie^6}{\Delta_M^2x}\bigg[G_\text{M}^*(\Delta_M^2x)\,g^{\sigma\tau}-{G_2^*(\Delta_M^2x)}\frac{p_1^\sigma p_2^\tau}{M_\Sigma\hat M}\bigg]\\
&\times\int\hskip-1mm\frac{\diff^4l}{(2\pi)^4}\,\frac{
\big(\tilde{B}_{\mu\nu\sigma}^{\Sigma^0}+\tilde{B}_{\mu\nu\sigma}^{\Lambda}\big)
\big(\tilde{L}_{\;\;\;\tau}^{\mu\nu}(q_1,q_2)-\tilde{L}_{\;\;\;\tau}^{\mu\nu}(q_2,q_1)\big)}
{(l-p_1)^2(l-p_2)^2}
\bigg\}\,.
\end{split}
\end{equation}
It is apparent that the contribution of the four terms arising in the numerator of the integrand above can be reconstructed from a single common term
\begin{equation}
\text{Tr}_{\sigma\tau}^{\text{BL},X}(q_1,q_2)
\equiv
\int\hskip-1mm\frac{\diff^4l}{(2\pi)^4}\,\frac{
\tilde{B}_{\mu\nu\sigma}^X
\tilde{L}_{\;\;\;\tau}^{\mu\nu}(q_1,q_2)}
{(l-p_1)^2(l-p_2)^2}\,.
\label{eq:TrBLX}
\end{equation}
Here it is important to stress that the baryonic part is invariant under the $q_1\leftrightarrow q_2$ exchange.
The whole contribution then reads
\begin{equation}
\begin{split}
&\overline{|\mathcal{M}^\text{LO+NLO}|^2}\Big|_{1\gamma\text{IR}}\\
&=\operatorname{Re}\bigg\{
\frac{ie^6}{\Delta_M^2x}\bigg[G_\text{M}^*(\Delta_M^2x)\,g^{\sigma\tau}-{G_2^*(\Delta_M^2x)}\frac{p_1^\sigma p_2^\tau}{M_\Sigma\hat M}\bigg]\\
&\times\sum_{X\in\{\Lambda,\Sigma^0\}}
\text{Tr}_{\sigma\tau}^{\text{BL},X}(q_1,q_2)
-(q_1\leftrightarrow q_2)
\bigg\}\,.
\end{split}
\label{eq:Mvirtbox}
\end{equation}
The correction is then simply given as
\begin{equation}
\delta^{1\gamma\text{IR}}(x,y)
=\overline{|\mathcal{M}^\text{LO+NLO}|^2}\Big|_{1\gamma\text{IR}}\bigg\slash\overline{|\mathcal{M}^\text{LO}(x,y)|^2}\,.
\label{eq:d1gIR}
\end{equation}

If the form factors $G_1^{XY}(q^2)$ and $G_2^{XY}(q^2)$ appearing in the loop were considered to be independent of the transferred momentum, it might have been sufficient to take into account the expansion in the sense of \ref{app:betaT}.
However, it is convenient here to involve in the calculation the more sophisticated form factors discussed in \ref{app:FF}, which requires a different redistribution of the terms based on the different behavior of $G_1^{XY}(q^2)$ and $G_2^{XY}(q^2)$.
The result for a particular model is then obtained by means of inserting the model-dependent linear combination
\begin{equation}
\begin{split}
\text{Tr}_{\sigma\tau}^{\text{BL},X}(q_1,q_2)
&=\sum_{i,j=1}^2
\sum_{k=1}^{N_i}
\sum_{l=1}^{N_j}
\bigg\{
c_i^{X\Lambda} c_j^{\Sigma^0X}
\alpha_{i,k}\alpha_{j,l}\\
&\times\text{Tr}_{\sigma\tau}^{\text{BL}(ij),X}(M_{i,k}^2,M_{j,l}^2;q_1,q_2)\bigg\}
\end{split}
\label{eq:TrBLXexp}
\end{equation}
into the prescription (\ref{eq:Mvirtbox}).
The building block $\text{Tr}_{\sigma\tau}^{\text{BL}(ij),X}(M_{i,k}^2,M_{j,l}^2;q_1,q_2)$ is calculated in \ref{app:BL}.
The pairs of coefficients $\alpha_{i,k}\equiv\alpha_{i,k}(\delta m^2)$ and $M_{i,k}\equiv M_{i,k}(\delta m^2)$ are, in the case of the model discussed in \ref{app:FF}, given by the following expansions:
\begin{equation}
\begin{split}
&\frac{q^2M_V^4}{(q^2-M_V^2)^3}\frac1{q^2}
=\lim_{\delta m^2\to0}\sum_{k=1}^{N_1}\frac{\alpha_{1,k}(\delta m^2)}{q^2-M_{1,k}^2(\delta m^2)}\\
&=\lim_{\delta m^2\to0}\frac{M_V^4}{2(\delta m^2)^2}
\left\{-\frac2{q^2-M_V^2}\right.\\
&+\frac1{q^2-(M_V^2+\delta m^2)}
+\frac1{q^2-(M_V^2-\delta m^2)}
\bigg\}\,,
\end{split}
\end{equation}
and
\begin{equation}
\begin{split}
&\frac{M_V^6}{(q^2-M_V^2)^3}\frac1{q^2}
=\lim_{\delta m^2\to0}\sum_{k=1}^{N_2}\frac{\alpha_{2,k}(\delta m^2)}{q^2-M_{2,k}^2(\delta m^2)}\\
&=-\frac1{q^2}+\lim_{\delta m^2\to0}\frac{M_V^4}{2(\delta m^2)^2}
\left\{-\frac2{q^2-M_V^2}\right.\\
&+\frac{M_V^2}{(M_V^2+\delta m^2)}\frac1{[q^2-(M_V^2+\delta m^2)]}+(\delta m^2\to-\delta m^2)
\bigg\}\,.
\end{split}
\end{equation}
The final result is then a lengthy linear combination of tensorial integrals defined in \ref{app:LI}.
Note that for the constant form factors $G_\text{E}$ and $G_\text{M}$ we would put simply $N_1=1$, $\alpha_{1,1}=1$, $M_{1,1}=M_V$ and $N_2=2$, $\alpha_{2,1}=1$, $\alpha_{2,2}=-1$, $M_{2,1}=M_V$, $M_{2,2}=0$.

\section{Correction to the \texorpdfstring{$\Sigma^0\Lambda\gamma$}{Sigma0-Lambda-gamma} vertex}
\label{sec:SLg}

In this section we would like to see if there are any significant electromagnetic corrections to the $\Sigma^0\Lambda\gamma$ vertex.
Due to its Lorentz structure we can write (cf.\ Eq.~(\ref{eq:SL_Lorentz}))
\begin{equation}
i\mathcal{M}_\mu^{\Sigma^0\Lambda\gamma^*}
=(ie)\bar u_\Lambda(\vec p_2)
\Gamma_\mu(p_1,p_2)
u_\Sigma(\vec p_1)\,,
\end{equation}
where $\Gamma_\mu$ now incorporates all the contributions in the QED expansion with the LO contribution fixed as
\begin{equation}
\Gamma_\mu^\text{LO}(p_1,p_2)
=G_\mu^{\Sigma^0\Lambda}(p_1-p_2)\,.
\end{equation}
At NLO, the correction is represented by four diagrams shown in Fig.~\ref{fig:virt_bar}:
\begin{equation}
\Gamma_\mu^\text{NLO}
=\sum_{X,Y}\Gamma_\mu^{XY}
=\Gamma_\mu^{\Sigma^0\Lambda}+\Gamma_\mu^{\Sigma^0\Sigma^0}+\Gamma_\mu^{\Lambda\Lambda}+\Gamma_\mu^{\Lambda\Sigma^0}\,.
\end{equation}
The building blocks $\Gamma_\mu^{XY}$ (with $X,Y\in\{\Lambda,\Sigma^0\}$) can then be written as (for brevity we drop the `$+i\epsilon$' parts of the propagators)
\begin{equation}
\begin{split}
&\Gamma_\mu^{XY}(p_1,p_2)\\
&=-i^5e^2\hskip-1mm\int\hskip-1mm\frac{\diff^4l}{(2\pi)^4}
\,\frac{G_\alpha^{Y\Lambda}(l)H_\mu^{XY}(l,p_1,p_2)G_\beta^{\Sigma^0X}(-l)g^{\alpha\beta}}
{[l^2][(l+p_1)^2-M_X^2][(l+p_2)^2-M_Y^2]}\,,
\end{split}
\label{eq:GmuXY}
\end{equation}
where
\begin{equation}
H_\mu^{XY}(l,p_1,p_2)
=(\slashed l+\slashed p_2+M_Y)G_\mu^{XY}(p_1-p_2)(\slashed l+\slashed p_1+M_X)\,.
\end{equation}
The off-shell form factors $G_\mu^{XY}$ are defined in Eq.~(\ref{eq:Goff}).

The magnetic-moment nature of the $\Sigma^0\Lambda\gamma$ interaction (its structure together with the fact that $G_1(l^2)\sim l^2$ when $l^2\to0$ and that $G_2(l^2)$ comes with $l^\rho$) prevents the appearance of the IR divergence, which could arise for $\{X,Y\}=\{\Sigma^0,\Lambda\}$.
On the other hand, due to the (loop-momenta-)power counting, $\Gamma_\mu^\text{NLO}$ is divergent in the UV domain (for the contribution proportional to $G_1^2(l^2)$), if the constant form factors (see Eq.~\eqref{eq:GEM_const}) are used.
That is why it is necessary to use a model with a stronger UV suppression, e.g.\ (\ref{eq:GEM_nonconst}).

Considering the loop integral (\ref{eq:GmuXY}), within the above mentioned model (\ref{eq:GEM_nonconst}), $G_\alpha^{Y\Lambda}(l)$ and $G_\beta^{\Sigma^0X}(-l)$ combine together through their components $G_i^{XY}((\pm l)^2)$ into the sixth power of the vector-meson propagator, which can be written as a fifth derivative with respect to the vector-meson mass:
\begin{equation}
\frac{(M_V^2)^n}{(l^2-M_V^2)^6}
=\frac1{5!}(M_V^2)^n\frac{\partial^5}{\partial(M_V^2)^5}\bigg[\frac1{l^2-M_V^2}\bigg]\,.
\end{equation}
This is particularly useful when one tries to obtain an analytic result.
For the purely numerical purpose, the approach described at the end of Section~\ref{sec:1gIR} is more suitable.

Numerically, the correction arising from the diagrams from Fig.~\ref{fig:virt_bar} is, as expected, negligible, and we show the results only for completeness.
It is probably interesting to see how this correction directly affects the form factor parameters and how stable are these in view of extraction from the experiment with respect to the QED corrections to the $\Sigma^0\Lambda\gamma$ vertex.
We present the complete numerical results and also the analytic ones for the NLO corrections to $G_2(q^2)$ as well as to $G_1(q^2)$ in the limit $M_\Sigma=\hat M=M_\Lambda$ (i.e.\ $\rho=0$) and for small $q^2$ (it is sufficient to work with a linear expansion in $q^2$).
Using the ratio
\begin{equation}
\sigma
\equiv\frac{M_V^2}{4\hat M^2}\,,
\end{equation}
we find for the Pauli form factor
\begin{equation}
G_2^\text{NLO}(q^2)
\simeq\frac\alpha\pi
\sum_{X,Y\in\{\Lambda,\Sigma^0\}}\,
\sum_{i,j=1}^2
G_\text{M}^{XY}(q^2)
\,c_i^{\Sigma^0X}c_j^{Y\Lambda}\hat Q_{XY}^{(ij)}(\sigma,\rho)\,,
\label{eq:G2NLO}
\end{equation}
where the functions $\hat Q_{XY}^{(ij)}(\sigma,\rho)$ can be decomposed in their real and imaginary parts separating the $\sigma$ and $\rho$ dependence in the following way:
\begin{equation}
\hat Q_{XY}^{(ij)}(\sigma,\rho)
\equiv Q^{(ij)}(\sigma)+iQ_{XY}^{(ij)}(\rho)\,.
\label{eq:Qhat}
\end{equation}
The real parts can be further decomposed as
\begin{equation}
\begin{split}
&Q^{(ij)}(\sigma)\\
&=\frac{\sigma^{2+i+j}}{5!}\frac{\diff^5}{\diff\sigma^5}
\left[\sigma P_1^{(ij)}(\sigma)\frac{\text{acos}(\sqrt{\sigma})}{\sqrt{1-\sigma}\sqrt{\sigma}}
+P_2^{(ij)}(\sigma)\log(\sigma)\right].
\end{split}
\label{eq:Qij}
\end{equation}
Note that for simplicity the terms vanishing after the derivatives are performed are not shown and that in our notation $G_\text{M}=G_\text{M}^\text{LO}=G_\text{M}^{\Sigma^0\Lambda}=G_\text{M}^{\Lambda\Sigma^0}$.
For the polynomials $P_k^{(ij)}(\sigma)=P_k^{(ji)}(\sigma)$ it holds:
\begin{align}
P_1^{(11)}(\sigma)&=-\sigma(3-4\sigma)\,,\quad P_2^{(11)}(\sigma)=\frac14\sigma(3-8\sigma)\,,\\
P_1^{(12)}(\sigma)&=-\sigma(2-3\sigma)\,,\quad P_2^{(12)}(\sigma)=\frac14\sigma(1-6\sigma)\,,\\
P_1^{(22)}(\sigma)&=2-5\sigma+4\sigma^2\,,\;\, P_2^{(22)}(\sigma)=-\frac12(1-3\sigma+4\sigma^2)\,.
\end{align}
In Eq.~(\ref{eq:G2NLO}) we also neglected the contribution of the electric form factor $G_\text{E}(q^2)$ and terms like $G_\text{M}(0)\,q^2/{\hat M^2}$.
The contributions to the imaginary part of $G_2^\text{NLO}(0)$ arise only for the combinations $\{X,Y\}=\{\Lambda,Y\}$ and, of course, need to be evaluated outside the $\rho\to0$ limit where they are non-vanishing.
Since the expressions are particularly simple, we show them here:
\begin{align}
&Q_{\Lambda\Lambda}^{(22)}(\rho)
=-\frac\pi4{(1-\sqrt{\rho})}\bigg[
\frac{1+\rho}{(1-\rho)^2}-\frac{\operatorname{atanh}(\sqrt{\rho})}{\sqrt{\rho}}
\bigg]\,,\\
\begin{split}
&Q_{\Lambda\Sigma^0}^{(22)}(\rho)
=-\frac\pi2(1+\rho)\\
&\times\left[1-\frac{\sqrt{\rho}}{{(1+\sqrt{\rho})^2}}
-\frac{1+\rho}{2{\sqrt{\rho}}}\log\bigg(\frac{(1+\sqrt{\rho})^2}{{1+\rho}}\bigg)\right].
\end{split}
\end{align}
Numerically, $Q_{\Lambda\Lambda}^{(22)}(\rho)=-2.255\times10^{-3}$ and $Q_{\Lambda\Sigma^0}^{(22)}(\rho)=-1.055\times10^{-3}$.
Other numerical results are compared in Tab.~\ref{tab:virt_num}.
We see how the numbers for respective contributions $\hat Q_{XY}^{(ij)}(\sigma,\rho)$ oscillate around the common approximate analytic result $Q^{(ij)}(\sigma)$.

\begin{table}[t!]
\centering
\small
\caption{
Values of the form-factor parameters used in the numerical evaluations.
Parameters $\kappa=G_\text{M}(0)$ are related to magnetic moments of particles of mass $M$ via $\mu=\kappa\frac e{2M}$; for the $\Sigma^0\Lambda$ transition we put $M=\hat M$.
The magnetic moments $\mu$ --- taken from Ref.~\cite{Kubis:2000aa} --- are expressed with respect to the nuclear magneton $\mu_\text{N}$.
Consequently, $\kappa=\frac{\mu}{\mu_\text{N}}\frac{M}{M_\text{p}}$.
The radii $\langle r_\text{M}^2\rangle$ and $\langle r_\text{E}^2\rangle$ are calculated based on the values from Ref.~\cite{Kubis:2000aa} and using $1\,\text{fm}=(1/0.197327)\,\text{GeV}^{-1}$.
Coefficients $c_1^{XY}$ are calculated using Eq.~(\ref{eq:c1M}) with $M_V=M_\rho$.
\label{tab:num}
}
\setlength{\tabcolsep}{1.5mm}
\begin{tabular}{c | c c c c || c }
\toprule
\noalign{\smallskip}
$XY$ & $\frac{\mu}{\mu_\text{N}}$ & $\kappa=-c_2^{XY}$ & $\frac{\langle r_\text{M}^2\rangle}{\text{GeV}^{-2}}$ & $\frac{\langle r_\text{E}^2\rangle}{\text{GeV}^{-2}}$ & $c_1^{XY}$\\
\noalign{\smallskip}
\midrule\midrule
\noalign{\smallskip}
$\Sigma^0\Lambda$
& $1.61(1)$ & $1.98(1)$ & $18.5(2.6)$ & $0.77(26)$ & $2.2(6)$\\
$\Sigma^0$
& $0.649$ & $0.825$ & $11.6(2.1)$ & $-0.77(26)$ & $1.5(2)$\\
$\Lambda$
& $-0.613$ & $-0.729$ & $12.3(2.3)$ & $2.8(5)$ & $-1.3(2)$\\
\noalign{\smallskip}
\bottomrule
\end{tabular}
\end{table}

\begin{table}[t!]
\centering
\small
\caption{
Analytic result $Q^{(ij)}(\sigma)$ compared to the numerical result for $\hat Q_{XY}^{(ij)}(\sigma,\rho)$ evaluated beyond the $\rho\to0$ limit, where, of course, the real part gains the dependence on $X$ and $Y$.
The values are to be multiplied by $10^{-3}$.
\label{tab:virt_num}
}
\begin{tabular}{c c | c c c c}
\toprule
\noalign{\smallskip}
$X$ & $Y$ & $(ij)=(11)$ & (12) & (21) & (22)\\
\noalign{\smallskip}
\midrule\midrule
\noalign{\smallskip}
\multicolumn{2}{c|}{$Q^{(ij)}(\sigma)$} & $-24.8$ & $-0.507$ & $-0.507$ & $-8.80$ \\
\noalign{\smallskip}\hline\noalign{\smallskip}
$\Sigma^0$ & $\Lambda$ & $-24.8$ & $-0.551$ & $-0.468$ & $-8.82$ \\
$\Sigma^0$ & $\Sigma^0$ & $-20.9$ & $-0.368$ & $-0.410$ & $-6.21$ \\
$\Lambda$ & $\Lambda$ & $-30.2$ & $-0.642$ & $-0.709$ & $-13.9-2.26i$ \\
$\Lambda$ & $\Sigma^0$ & $-25.3$ & $-0.517$ & $-0.517$ & $-9.81-1.06i$ \\
\noalign{\smallskip}
\bottomrule
\end{tabular}
\end{table}

The full result based on Eq.~(\ref{eq:G2NLO}) and values from Tab.~\ref{tab:num} numerically reads $\operatorname{Re}{G_2^\text{NLO}(q^2)}\allowbreak=\frac\alpha\pi(-0.36(16)-0.95(44)\frac{q^2}{\text{GeV}^2})$.
This is negligible in the global context, but is of the same order as Schwinger's correction to the magnetic moment of the electron~\cite{Schwinger:1949ra}.

Considering the numerical values, $\kappa_\Lambda\approx-\kappa_{\Sigma^0}$ and $\langle r_\text{M}^2\rangle_{\Sigma^0}\approx\langle r_\text{M}^2\rangle_\Lambda$; cf.\ Tab.~\ref{tab:num}.
Consequently, in the model (\ref{eq:GEM_nonconst}), $c_2^{\Sigma^0\Sigma^0}\approx-c_2^{\Lambda\Lambda}$ and $c_1^{\Lambda\Lambda}\approx-c_1^{\Sigma^0\Sigma^0}$.
Also, more generally and less precisely, $c_1\approx-c_2$.
These considerations lead to the fact that Eq.~(\ref{eq:G2NLO}) can be put into a form in which the form-factor parameters decouple:
\begin{equation}
\begin{split}
&\operatorname{Re}G_2^\text{NLO}(q^2)
\\
&\simeq-\frac\alpha\pi
\sum_{X,Y\in\{\Lambda,\Sigma^0\}}
G_\text{M}^{XY}(q^2)
\,c_1^{\Sigma^0X}c_2^{Y\Lambda}\sum_{i,j=1}^2(-1)^{i+j}Q^{(ij)}(\sigma)\,.
\end{split}
\label{eq:G2NLOb}
\end{equation}
Note that since $|Q^{(ij)}(\sigma)|\ll|Q^{(ii)}(\sigma)|,\,i\ne j$ (see Tab.~\ref{tab:virt_num}) the second sum in Eq.~(\ref{eq:G2NLOb}) reduces simply to $Q^{(11)}(\sigma)+Q^{(22)}(\sigma)$.
Finally, note that $Q^{(11)}(\sigma)$ has a nonzero limit as $\sigma\to0$ (coming from the coefficient of the linear term of $P_2^{11}(\sigma)$ times $(-3!)/5!$): $Q^{(11)}(0)=-3/80=-0.0375$.
From this fact one could infer that the contribution from $Q^{(11)}(\sigma)$ will be the most important one, as one can see indeed in Tab.~\ref{tab:virt_num}.

\begin{table}[t!]
\centering
\small
\caption{
Analytic result $Q_\text{M}^{(ij)}(\sigma)$ and $Q_\text{E}^{(ij)}(\sigma)$ compared to the numerical result evaluated beyond the $\rho\to0$ limit.
The values are to be multiplied by $10^{-3}$.
\label{tab:virt_num_1}
}
\begin{tabular}{c c | c c c c}
\toprule
\noalign{\smallskip}
$X$ & $Y$ & $(ij)=(11)$ & (12) & (21) & (22)\\
\noalign{\smallskip}
\midrule
\midrule
\noalign{\smallskip}
\multicolumn{2}{c|}{$Q_\text{M}^{(ij)}(\sigma)$} & $28.3$ & $7.65$ & $7.65$ & $4.37$ \\
\noalign{\smallskip}\hline\noalign{\smallskip}
$\Sigma^0$ & $\Lambda$ & $28.3$ & $7.85$ & $7.49$ & $4.37$ \\
$\Sigma^0$ & $\Sigma^0$ & $25.0$ & $6.47$ & $6.31$ & $1.84$ \\
$\Lambda$ & $\Lambda$ & $32.6$ & $9.51$ & $9.26$ & $9.56+2.33i$ \\
$\Lambda$ & $\Sigma^0$ & $29.0$ & $7.83$ & $7.83$ & $4.78+0.849i$ \\
\noalign{\smallskip}
\midrule\midrule
\noalign{\smallskip}
\multicolumn{2}{c|}{$Q_\text{E}^{(ij)}(\sigma)$} & $-21.5$ & $-6.88$ & $-6.88$ & $26.0$ \\
\noalign{\smallskip}\hline\noalign{\smallskip}
$\Sigma^0$ & $\Lambda$ & $-21.5$ & $-6.96$ & $-6.82$ & $26.1$ \\
$\Sigma^0$ & $\Sigma^0$ & $-18.5$ & $-5.75$ & $-5.68$ & $18.9$ \\
$\Lambda$ & $\Lambda$ & $-25.6$ & $-8.53$ & $-8.42$ & $39.9+6.55i$ \\
$\Lambda$ & $\Sigma^0$ & $-22.0$ & $-7.03$ & $-7.03$ & $29.0+3.16i$ \\
\noalign{\smallskip}
\bottomrule
\end{tabular}
\end{table}

Along similar lines, one can calculate the corrections to the Dirac form factor.
In this case, the electric form factor should not be neglected from the beginning, since the contribution proportional to it is possibly of a similar size as the one stemming from the magnetic form factor.
We find
\begin{equation}
G_1^\text{NLO}(q^2)
\simeq\frac\alpha\pi
\sum_{I}
\,
\sum_{X,Y}
H_I^{XY}(q^2)
\sum_{i,j=1}^2
c_i^{\Sigma^0X}c_j^{Y\Lambda}\hat Q_{I,XY}^{(ij)}(\sigma,\rho)
\,,
\label{eq:G1NLO}
\end{equation}
with $I\in\{\text{E},\text{M}\}$,
\begin{align}
H_\text{E}^{XY}(q^2)&=G_\text{E}^{XY}(q^2)\,,\\
H_\text{M}^{XY}(q^2)&=\frac{q^2}{4\hat M^2}G_\text{M}^{XY}(q^2)\,,
\end{align}
and, analogically to Eq.~(\ref{eq:Qhat}),
\begin{equation}
\hat Q_{I,XY}^{(ij)}(\sigma,\rho)
\equiv Q_I^{(ij)}(\sigma)+iQ_{I,XY}^{(ij)}(\rho)\,.
\end{equation}
The real parts $Q_I^{(ij)}(\sigma)$ can be decomposed in terms of polynomials $P_{I,k}^{(ij)}(\sigma)=P_{I,k}^{(ji)}(\sigma)$ in the same way as it is shown in Eq.~(\ref{eq:Qij}), with (for brevity we drop denoting explicitly the $\sigma$ dependence of the polynomials)
\begin{flalign}
P_{\text{E},1}^{(11)}&=-3\sigma(1+2\sigma-4\sigma^2)\,,
\hspace{-2mm}
&P_{\text{M},1}^{(11)}&=\sigma[1+2\sigma(5-6\sigma)]\,,
\\ 
P_{\text{E},2}^{(11)}&=\frac34\sigma(1-8\sigma^2)\,,
&&\hspace{-15mm}P_{\text{M},2}^{(11)}=-\frac14\sigma[3+8\sigma(1-3\sigma)]\,,
\\
P_{\text{E},1}^{(12)}&=-3\sigma(4-5\sigma)\,,
&P_{\text{M},1}^{(12)}&=\sigma(10-11\sigma)\,,
\\
P_{\text{E},2}^{(12)}&=\frac34\sigma(3-10\sigma)\,,
&P_{\text{M},2}^{(12)}&=-\frac14\sigma(9-22\sigma)\,,
\\
P_{\text{E},1}^{(22)}&=-3(2-\sigma-2\sigma^2)\,,
&P_{\text{M},1}^{(22)}&=-6+19\sigma-14\sigma^2\,,
\\
P_{\text{E},2}^{(22)}&=\frac32(1-2\sigma-2\sigma^2)\,,
&P_{\text{M},2}^{(22)}&=\frac12(1-12\sigma+14\sigma^2)\,.
\end{flalign}

Once more, for completeness, we also show the analytic forms of the contributions to the imaginary part of $G_1^\text{NLO}(q^2)$:
\begin{align}
Q_{\text{E},\Lambda\Lambda}^{(22)}(\rho)
&=\frac{2\pi\rho}{(1-\rho)\left(1+\sqrt{\rho}\right)^2}\,,\\
Q_{\text{M},\Lambda\Lambda}^{(22)}(\rho)
&=\frac{\pi}{4}\left[\frac{1-3\rho}{\left(1+\sqrt{\rho}\right)^3}+\left(3-\frac{1}{\sqrt{\rho}}\right)\operatorname{atanh}\left(\sqrt{\rho}\right)\right],\\
Q_{\text{E},\Lambda\Sigma^0}^{(22)}(\rho)
&=\pi\rho\left[\frac{1-\sqrt{\rho}-\rho}{\left(1+\sqrt{\rho}\right)^2}+\frac{\sqrt{\rho}}{2}\log\left(\frac{\left(1+\sqrt{\rho}\right)^2}{1+\rho}\right)\right],\\
\begin{split}
Q_{\text{M},\Lambda\Sigma^0}^{(22)}(\rho)
&=\frac{\pi}{2}\left[\frac{1+3\rho}{\left(1+\sqrt{\rho}\right)^3}+\frac{(2+5\rho)\sqrt{\rho}}{1+\sqrt{\rho}}\right.\\
&-\left.\frac{(1+\rho)(1+5\rho)}{2\sqrt{\rho}}\log\left(\frac{\left(1+\sqrt{\rho}\right)^2}{1+\rho}\right)\right].
\end{split}
\end{align}
Numerically, $Q_{\text{E},\Lambda\Lambda}^{(22)}(\rho)=6.548\times10^{-3}$, $Q_{\text{M},\Lambda\Lambda}^{(22)}(\rho)=2.325\times10^{-3}$, $Q_{\text{E},\Lambda\Sigma^0}^{(22)}(\rho)=3.161\times10^{-3}$ and $Q_{\text{M},\Lambda\Sigma^0}^{(22)}(\rho)=0.8489\times10^{-3}$.
Other numerical results are compared in Tab.~\ref{tab:virt_num_1}.
For the full correction based on Eq.~(\ref{eq:G1NLO}) and values from Tab.~\ref{tab:num} we find $\operatorname{Re}G_1^\text{NLO}(q^2)\allowbreak=\frac\alpha\pi0.040(38)\frac{q^2}{\text{GeV}^2}$.

Similarly as we derived Eq.~(\ref{eq:G2NLOb}), we can find an approximate formula for $\operatorname{Re}G_1^\text{NLO}(q^2)$, which is obtained from Eq.~(\ref{eq:G2NLOb}) using the substitution $Q^{(ij)}(\sigma)\to Q_\text{M}^{(ij)}(\sigma)\,q^2/(4\hat M^2)$.

Let us now see how the corrections to the Dirac and Pauli form factors translate into the NLO corrections $\Delta X\equiv X^\text{NLO}$ to the parameters $X^\text{LO}$ of the electric and magnetic form factors in the linear expansions of Eqs.\ \eqref{eq:expandGEGM}.
We find
\begin{align}
\Delta\kappa
&=\Delta G_\text{M}(0)=G_2^\text{NLO}(0)\,,\label{eq:Dkappa}\\
\Delta\langle r_\text{E}^2\rangle
&=6\left[\frac{\diff G_1^\text{NLO}(q^2)}{\diff q^2}\bigg|_{q^2=0}
+\frac{G_2^\text{NLO}(0)}{4\hat M^2}\right]\,,\label{eq:DrE2}\\
\Delta\langle r_\text{M}^2\rangle
&=\frac6\kappa\frac{\diff G_\text{M}^\text{NLO}(q^2)}{\diff q^2}\bigg|_{q^2=0}
-\frac{\Delta\kappa}\kappa\langle r_\text{M}^2\rangle
\label{eq:DrM2},
\end{align}
with $G_\text{M}^\text{NLO}(q^2)=G_1^\text{NLO}(q^2)+G_2^\text{NLO}(q^2)$.
Note that strictly speaking, in our notation, all the parameters here could wear the $\Sigma^0\Lambda$ superscripts.
Numerically, the relative corrections $\delta_X\equiv\operatorname{Re}\Delta X/X$ are negligible: based on values from Tab.~\ref{tab:num}, $\delta_{\kappa}=-0.042(19)\,\%$, $\delta_{\langle r_\text{E}^2\rangle}=-(0.050_{-27}^{+61})\,\%$ and $\delta_{\langle r_\text{M}^2\rangle}=0.0071_{-26}^{+40}\,\%$; the numerical cancellations among the terms in Eq.~(\ref{eq:DrM2}) are responsible for the smallness of the last correction.
Finally, the correction to the differential decay width takes the form
\begin{equation}
\begin{split}
\delta_{\Sigma^0\Lambda\gamma}^\text{virt}(x,y)
&\equiv\frac{\overline{|\mathcal{M}_G^\text{LO+NLO}(x,y)|^2}}{\overline{|\mathcal{M}_G^\text{LO}(x,y)|^2}}-1\\
&\simeq2\operatorname{Re}\bigg\{\frac{\Delta\kappa}\kappa+\frac16\Delta\langle r_\text{M}^2\rangle x\Delta_M^2\bigg\}\,.
\end{split}
\end{equation}
Above, $\overline{|\mathcal{M}_G^\text{LO}(x,y)|^2}$ was defined in Eq.~(\ref{eq:MGLOxy}) and $\overline{|\mathcal{M}_G^\text{LO+NLO}(x,y)|^2}$ is its equivalent with form factors taken up to NLO in the view of this section (with the parameters corrected according to Eqs.~(\ref{eq:Dkappa}-\ref{eq:DrM2})).
Up to the linear order in $x$, there is no $y$-dependence.
On top of that, the $x$-dependence is very weak due to the numerical insignificance of the linear term ($\frac26\Delta_M^2\operatorname{Re}\Delta\langle r_\text{M}^2\rangle\sim10^{-6}$) within the kinematically allowed region.
Consequently, $\delta_{\Sigma^0\Lambda\gamma}^\text{virt}(x,y)\simeq\delta_{\Sigma^0\Lambda\gamma}^\text{virt}\simeq2\delta_\kappa=-0.084(38)\,\%$.
The total correction due to the QED NLO effects on the $\Sigma^0\Lambda\gamma$ vertex thus does not effect the measurement of the slope parameters (the electric and magnetic radii) and has only a tiny effect on the magnetic moment $\kappa$ and thereupon on the decay rate.

\section{Results and discussion}
\label{sec:res}

With all the calculations at hand, we can now answer the following questions:
\begin{enumerate}
\item Suppose one measures the differential decay width for
$\Sigma^0\to(\Lambda e^+e^- +\,$arbitrary many photons).
How large are the radiative corrections which relate this measurement to the QED leading-order calculation for $\Sigma^0\to\Lambda e^+e^-$?
In particular, which aspects of the hadronic transition form factors can be extracted from such measurements and how can they be extracted?
\item Can one safely neglect those radiative corrections that involve other hadronic form factors, i.e.\ the diagrams depicted
in Figs.\ \ref{fig:1gIR}, \ref{fig:1gIR2}, and \ref{fig:virt_bar}?
\item What are the differences between the calculations presented here and the ones in Ref.~\cite{Sidhu:1972rx}?
\end{enumerate}
We first provide a quick overview of the first two questions and then discuss the results in detail.
Finally we will come back to the third question.

\begin{figure}[!t]
\centering
\includegraphics[width=1.\columnwidth]{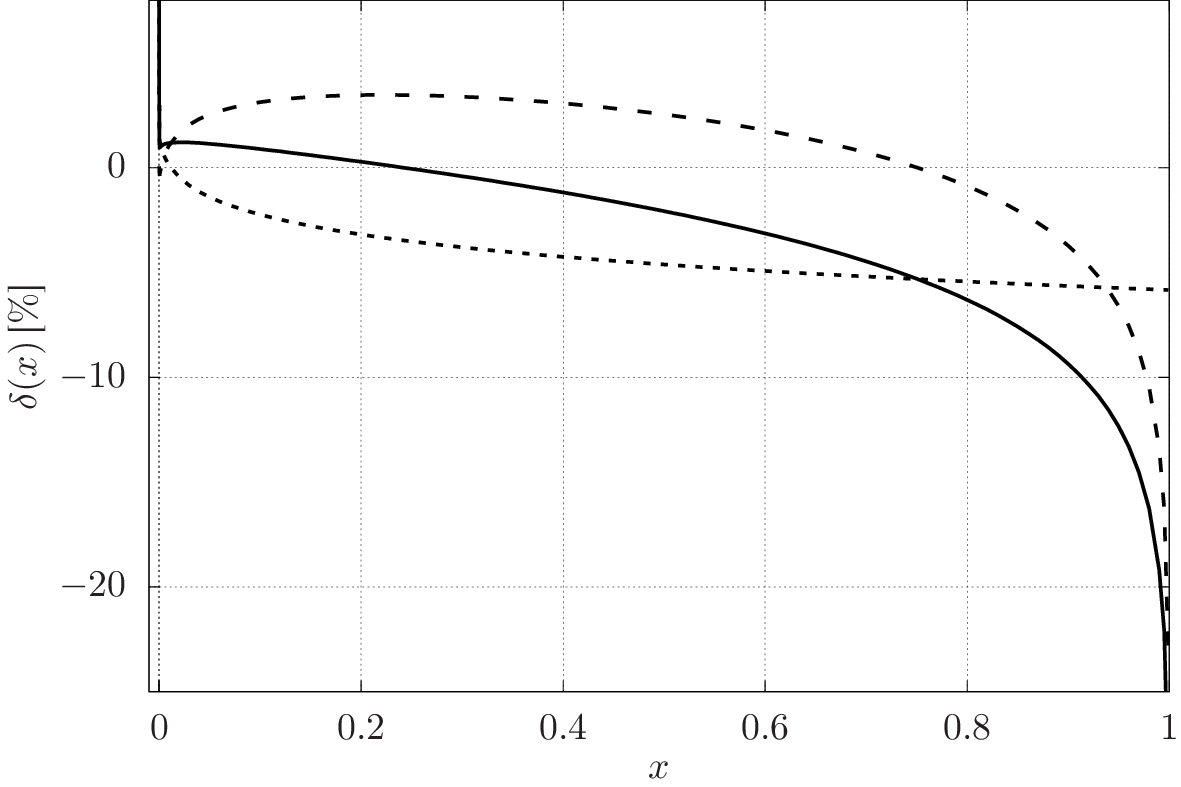}
\caption{\label{fig:deltax}
The total NLO correction $\delta(x)$ for the decay $\Sigma^0\to\Lambda e^+e^-$ (solid line) in comparison to its constituents.
The virtual correction $\delta^\text{virt}(x)$ is depicted as a dotted line.
The bremsstrahlung correction $\delta^\text{BS}(x)$ is shown as a dashed line.
The divergent behavior of $\delta(x)$ near $x=\nu^2\approx0$ has the origin in the electromagnetic form factor $F_1(x)$ and is connected to the Coulomb self-interaction of the dilepton at the threshold.
This divergence is integrable.
}
\end{figure}

The short answer to the second question is yes.
To answer the first question comprehensively, we distinguish the two-fold differential, one-fold differential and integrated decay rate.
Following the definitions of Eq.~\eqref{eq:dxydx} we provide the relative changes for the differential distributions.
The corresponding results are shown in Tab.~\ref{tab:deltaSym} and Fig.~\ref{fig:deltax}, respectively.

\begin{table*}[!t]
\centering
\scriptsize
\caption{
The NLO correction $\delta(x,y)-\delta^{1\gamma\text{IR}}(x,y)$ (i.e.\ excluding the 1$\gamma$IR contribution) given {\em in percent} for a range of values $x$ and $y$ (i.e.\ the Dalitz-plot corrections) for the process $\Sigma^0\to\Lambda e^+e^-$.
It is sufficient to show the results for positive values of $y$ only since these corrections are {\em symmetric} under $y\to-y$.
The larger values at the edge of the kinematically allowed region (as $x\to1$) are naturally present due to the fact that the correction itself is defined as a ratio of the NLO and LO decay widths which both vanish for $x\to1$.
}
\label{tab:deltaSym}
\setlength{\tabcolsep}{4mm}
\begin{tabular}{c r r r r r r r r r r r}
\toprule
$x$ $\bigg\backslash$ $y$ & 0.00 & 0.10 & 0.20 & 0.30 & 0.40 & 0.50 & 0.60 & 0.70 & 0.80 & 0.90 & 0.99 \\
\midrule
\noalign{\smallskip}
0.01 & 2.50 & 2.44 & 2.31 & 2.17 & 2.03 & 1.87 & 1.65 & 1.33 & 0.83 & -0.20 & -8.26 \\
0.02 & 2.67 & 2.61 & 2.49 & 2.34 & 2.18 & 2.00 & 1.75 & 1.40 & 0.84 & -0.26 & -5.84 \\
0.03 & 2.71 & 2.66 & 2.55 & 2.41 & 2.24 & 2.04 & 1.78 & 1.41 & 0.83 & -0.33 & -5.75 \\
0.04 & 2.71 & 2.67 & 2.56 & 2.42 & 2.26 & 2.06 & 1.79 & 1.41 & 0.80 & -0.40 & -5.84 \\
0.05 & 2.69 & 2.65 & 2.55 & 2.42 & 2.26 & 2.05 & 1.78 & 1.39 & 0.77 & -0.47 & -5.98 \\
0.06 & 2.66 & 2.62 & 2.53 & 2.40 & 2.24 & 2.04 & 1.76 & 1.37 & 0.73 & -0.53 & -6.13 \\
0.07 & 2.61 & 2.58 & 2.49 & 2.37 & 2.21 & 2.01 & 1.74 & 1.34 & 0.69 & -0.60 & -6.29 \\
0.08 & 2.56 & 2.53 & 2.45 & 2.33 & 2.18 & 1.98 & 1.71 & 1.30 & 0.65 & -0.66 & -6.44 \\
0.09 & 2.51 & 2.48 & 2.41 & 2.29 & 2.15 & 1.95 & 1.67 & 1.27 & 0.60 & -0.73 & -6.60 \\
\\
0.10 & 2.45 & 2.43 & 2.36 & 2.25 & 2.11 & 1.91 & 1.64 & 1.23 & 0.56 & -0.79 & -6.75 \\
0.15 & 2.14 & 2.12 & 2.07 & 1.99 & 1.87 & 1.69 & 1.42 & 1.01 & 0.31 & -1.12 & -7.47 \\
0.20 & 1.79 & 1.78 & 1.75 & 1.69 & 1.59 & 1.43 & 1.17 & 0.75 & 0.04 & -1.46 & -8.14 \\
0.25 & 1.43 & 1.42 & 1.40 & 1.36 & 1.28 & 1.14 & 0.89 & 0.48 & -0.26 & -1.81 & -8.78 \\
0.30 & 1.05 & 1.05 & 1.04 & 1.01 & 0.95 & 0.82 & 0.59 & 0.17 & -0.57 & -2.18 & -9.40 \\
0.35 & 0.65 & 0.65 & 0.65 & 0.64 & 0.59 & 0.48 & 0.26 & -0.15 & -0.91 & -2.57 & -10.0\,\,\, \\
0.40 & 0.23 & 0.23 & 0.24 & 0.24 & 0.21 & 0.11 & -0.10 & -0.51 & -1.28 & -2.99 & -10.6\,\,\, \\
0.45 & -0.22 & -0.22 & -0.20 & -0.18 & -0.20 & -0.29 & -0.49 & -0.89 & -1.68 & -3.43 & -11.2\,\,\, \\
0.50 & -0.71 & -0.70 & -0.67 & -0.64 & -0.65 & -0.72 & -0.91 & -1.31 & -2.11 & -3.91 & -11.9\,\,\, \\
\\
0.55 & -1.23 & -1.22 & -1.18 & -1.15 & -1.14 & -1.20 & -1.38 & -1.78 & -2.59 & -4.43 & -12.6\,\,\, \\
0.60 & -1.81 & -1.79 & -1.75 & -1.70 & -1.68 & -1.73 & -1.90 & -2.30 & -3.12 & -5.01 & -13.3\,\,\, \\
0.65 & -2.45 & -2.44 & -2.38 & -2.32 & -2.29 & -2.32 & -2.49 & -2.89 & -3.72 & -5.65 & -14.1\,\,\, \\
0.70 & -3.19 & -3.16 & -3.10 & -3.03 & -2.98 & -3.01 & -3.17 & -3.56 & -4.41 & -6.38 & -14.9\,\,\, \\
0.75 & -4.04 & -4.01 & -3.94 & -3.86 & -3.80 & -3.81 & -3.96 & -4.36 & -5.22 & -7.23 & -15.9\,\,\, \\
0.80 & -5.06 & -5.03 & -4.96 & -4.86 & -4.79 & -4.79 & -4.93 & -5.33 & -6.21 & -8.26 & -17.0\,\,\, \\
0.85 & -6.36 & -6.33 & -6.24 & -6.14 & -6.05 & -6.04 & -6.18 & -6.58 & -7.47 & -9.56 & -18.4\,\,\, \\
0.90 & -8.16 & -8.12 & -8.03 & -7.91 & -7.81 & -7.79 & -7.92 & -8.32 & -9.24 & -11.4\,\,\, & -20.3\,\,\, \\
0.95 & -11.2\,\,\, & -11.1\,\,\, & -11.0\,\,\, & -10.9\,\,\, & -10.8\,\,\, & -10.8\,\,\, & -10.9\,\,\, & -11.3\,\,\, & -12.2\,\,\, & -14.4\,\,\, & -23.4\,\,\, \\
0.99 & -18.0\,\,\, & -18.0\,\,\, & -17.9\,\,\, & -17.7\,\,\, & -17.6\,\,\, & -17.6\,\,\, & -17.7\,\,\, & -18.1\,\,\, & -19.0\,\,\, & -21.2\,\,\, & -30.3\,\,\, \\
\noalign{\smallskip}
\bottomrule
\end{tabular}
\end{table*}

Concerning the integrated width for the $\Sigma^0\to\Lambda e^+e^-$ decay, it can be meaningful to normalize to the LO integrated decay width or to the rate of the real-photon decay of $\Sigma^0\to\Lambda\gamma$.
In the latter case one can simplify and systemize the result by neglecting the electric transition form factor and linearizing the magnetic one according to the discussion around Eq.~\eqref{eq:expandGEGM}.
Then the ratio
\begin{equation}
R
\equiv\frac{\Gamma(\Sigma^0\to\Lambda e^+e^-)}{\Gamma(\Sigma^0\to\Lambda\gamma)}
\label{eq:R}
\end{equation}
is independent of $G_{\text{M}}(0)$ and depends only on one hadronic quantity, the magnetic transition radius.
Consequently, one can write
\begin{equation}
R
=R_0+aR_1+\mathcal{O}(a^2)
\label{eq:Rsplitup-magrad}
\end{equation}
with
\begin{equation}
a\equiv\frac16\langle r_\text{M}^2\rangle\Delta_M^2\,.
\label{eq:defamagrad}
\end{equation}
Results are provided in Tab.~\ref{tab:R}.
After this brief summary we turn to the details.

In the previous sections, we discussed the main parts of the radiative corrections as we referred to them in Eq.~(\ref{eq:delta_origins}), i.e.\ the virtual corrections in Section~\ref{sec:virt} together with the related correction to the $\Sigma^0\Lambda\gamma$ vertex in Section \ref{sec:SLg}, the bremsstrahlung from the lepton legs evaluated beyond the soft-photon approximation in Section~\ref{sec:BS} and the 1$\gamma$IR correction in Section~\ref{sec:1gIR}.
In order to get the final correction, one then simply sums over the partial results found to be non-negligible: Eqs.~(\ref{eq:dvirtfinal}) and (\ref{eq:deltaBS}).
Let us now comment on these contributions in detail.

Considering the virtual corrections related to the photon and lepton lines and discussed in Section~\ref{sec:virt}, the correction to the one-fold differential decay width $\delta^\text{virt}(x)$ from Eq.~(\ref{eq:dvirtxfinal}) is model-independent.
To a very large extent this is also true for the correction to the two-fold differential decay width $\delta^\text{virt}(x,y)$ from Eq.~(\ref{eq:dvirtfinal}):
Firstly, the electric form factor can be safely neglected here, which makes the magnetic part cancel out.
Secondly, the numerical contribution of the $F_2^\text{NLO}(q^2)$ part (also containing the model dependence) is negligible compared to the rest of the expression.
The form factor $\overline{F_1^\text{NLO}}(\Delta_M^2x)$ then contains the IR-divergent piece which cancels with the corresponding term stemming from the bremsstrahlung contribution, as shown at the end of Section~\ref{sec:BS}.

Similarly, the bremsstrahlung contribution (\ref{eq:deltaBS}) discussed in Section~\ref{sec:BS} can be considered to be model-independent.
Firstly, it is evident that the divergent part of the bremsstrahlung matrix element squared (\ref{eq:MBSdiv}) leads to the model-independent correction; besides, this is governed by gauge invariance.
Secondly, after the electric form factor is neglected, the expansion (\ref{eq:GMBS}) allows for the cancellation of the magnetic form factor with the one appearing in the LO expression, leaving only terms dependent on its slope value in the final result for $\delta^\text{BS}(x,y)$.
All the necessary definitions are then listed in \ref{app:J}.

\begin{table}[!t]
\centering
\small
\caption{
Radiative corrections based on their origin and their respective contributions to the ratio \eqref{eq:Rsplitup-magrad} and to the total correction $\delta$ to the decay rate.
The first column shows $R$ at LO.
The subscripts `C' and `D' at the BS correction correspond to the (IR-)convergent and divergent parts, respectively.
The values of $R_i$ are to be multiplied by $10^{-3}$.
\label{tab:R}
}
\begin{tabular}{c | c c c c || c}
\toprule
\noalign{\smallskip}
 & LO & virt & BS$|_\text{C}$ & BS$|_\text{D}$ & total\\
\noalign{\smallskip}
\midrule\midrule
\noalign{\smallskip}
$R_0$
& $5.484$ & $-0.0167$ & $-0.06443$\;\;\, & $0.1302$ & $5.533$\\
$R_1$
& $0.619$ & $-0.0201$ & $0.00010$ & $0.0287$ & $0.628$\\
\noalign{\smallskip}\hline\noalign{\smallskip}
$\delta$\,[\%] & --- & $-0.310$ & $-1.17$ & $2.38$ & $0.896$\\
\noalign{\smallskip}
\bottomrule
\end{tabular}
\end{table}

Finally, for the model-dependent 1$\gamma$IR contribution (\ref{eq:d1gIR}), as presented in Section~\ref{sec:1gIR}, we used the model discussed in \ref{app:FF}.
Let us again stress at this point that we used a rather general approach applicable to a wide family of rational models.
The final result within a particular model can then be related to an appropriate linear combination (\ref{eq:TrBLXexp}) of the building blocks (\ref{eq:TrBLijX}) defined in \ref{app:BL}.
We observe, which is soothing, that different models would lead to compatible values in the numerical results:
We explicitly checked this using the two models (\ref{eq:GEM_const}) and (\ref{eq:GEM_nonconst}) discussed in \ref{app:FF}.
The numerical results for the 1$\gamma$IR correction $\delta^{1\gamma\text{IR}}(x,y)$ within the model (\ref{eq:GEM_nonconst}) and using the values from Tab.~\ref{tab:num} are shown in Tab.~\ref{tab:deltaAsym}.
We can see that this correction is negligible and would only make cosmetic changes if included in Tab.~\ref{tab:deltaSym}; note the overall relative factor of 0.01 between the values presented in Tabs.~\ref{tab:deltaSym} and \ref{tab:deltaAsym}.
Moreover, what is important is the antisymmetric nature of this correction with respect to the electron--positron exchange, i.e.\ with respect to the sign change of the kinematic variable $y$.
This means that the correction $\delta^{1\gamma\text{IR}}(x)$
is necessarily vanishing and has thus no effect on the measurement of the form-factor slope or the total decay rate.
Note that this is also consistent with the fact that the interference of the $\delta^{1\gamma\text{IR}}(x)$ diagrams (Figs.~\ref{fig:1gIR} and \ref{fig:1gIR2}) with the LO diagram from Fig.~\ref{fig:LO} gives a contribution to the imaginary part of a two-loop diagram vanishing due to Furry's theorem; see also Ref.~\cite{Fael:2016yle}.
Therefore, after the phase-space integral is performed, this interference should indeed vanish accordingly.

\begin{table*}[t]
\centering
\scriptsize
\caption{
The 1$\gamma$IR correction $\delta^{1\gamma\text{IR}}(x,y)$ at NLO for the process $\Sigma^0\to\Lambda e^+e^-$ to be multiplied by $10^{-4}$.
It is sufficient to show the results for positive values of $y$ only since these corrections are {\em antisymmetric} under $y\to-y$ (and thus $\delta^{1\gamma\text{IR}}(x,0)=0$).
Note that instead of values for $y=0$, the first column shows $\delta^{1\gamma\text{IR}}(x,0.05)$.
}
\label{tab:deltaAsym}
\setlength{\tabcolsep}{4mm}
\begin{tabular}{c r r r r r r r r r r r}
\toprule
$x$ $\bigg\backslash$ $y$ & 0.05 & 0.10 & 0.20 & 0.30 & 0.40 & 0.50 & 0.60 & 0.70 & 0.80 & 0.90 & 0.99 \\
\midrule
\noalign{\smallskip}
0.01 & -0.00 & -0.00 & -0.00 & -0.01 & -0.01 & -0.01 & -0.01 & -0.01 & -0.01 & -0.01 & -0.00 \\
0.02 & -0.00 & -0.01 & -0.01 & -0.01 & -0.02 & -0.02 & -0.02 & -0.02 & -0.02 & -0.01 & -0.01 \\
0.03 & -0.00 & -0.01 & -0.01 & -0.02 & -0.02 & -0.03 & -0.03 & -0.03 & -0.02 & -0.02 & -0.01 \\
0.04 & -0.01 & -0.01 & -0.02 & -0.03 & -0.03 & -0.03 & -0.03 & -0.03 & -0.03 & -0.03 & -0.02 \\
0.05 & -0.01 & -0.01 & -0.02 & -0.03 & -0.04 & -0.04 & -0.04 & -0.04 & -0.04 & -0.03 & -0.03 \\
0.06 & -0.01 & -0.01 & -0.03 & -0.04 & -0.05 & -0.05 & -0.05 & -0.05 & -0.05 & -0.04 & -0.04 \\
0.07 & -0.01 & -0.02 & -0.03 & -0.05 & -0.05 & -0.06 & -0.06 & -0.06 & -0.06 & -0.05 & -0.04 \\
0.08 & -0.01 & -0.02 & -0.04 & -0.05 & -0.06 & -0.07 & -0.07 & -0.07 & -0.07 & -0.06 & -0.05 \\
0.09 & -0.01 & -0.02 & -0.04 & -0.06 & -0.07 & -0.08 & -0.08 & -0.08 & -0.08 & -0.07 & -0.06 \\
\\
0.10 & -0.01 & -0.02 & -0.05 & -0.07 & -0.08 & -0.09 & -0.09 & -0.09 & -0.09 & -0.08 & -0.07 \\
0.15 & -0.02 & -0.04 & -0.07 & -0.10 & -0.13 & -0.14 & -0.15 & -0.15 & -0.14 & -0.13 & -0.12 \\
0.20 & -0.03 & -0.05 & -0.10 & -0.14 & -0.18 & -0.20 & -0.21 & -0.21 & -0.21 & -0.20 & -0.18 \\
0.25 & -0.04 & -0.07 & -0.13 & -0.19 & -0.23 & -0.26 & -0.28 & -0.29 & -0.28 & -0.27 & -0.25 \\
0.30 & -0.04 & -0.09 & -0.17 & -0.24 & -0.29 & -0.33 & -0.36 & -0.37 & -0.36 & -0.35 & -0.33 \\
0.35 & -0.05 & -0.11 & -0.21 & -0.29 & -0.36 & -0.41 & -0.44 & -0.46 & -0.46 & -0.44 & -0.42 \\
0.40 & -0.07 & -0.13 & -0.25 & -0.36 & -0.44 & -0.50 & -0.54 & -0.56 & -0.56 & -0.55 & -0.53 \\
0.45 & -0.08 & -0.15 & -0.30 & -0.42 & -0.52 & -0.60 & -0.65 & -0.67 & -0.68 & -0.67 & -0.65 \\
0.50 & -0.09 & -0.18 & -0.35 & -0.50 & -0.62 & -0.71 & -0.77 & -0.80 & -0.81 & -0.81 & -0.78 \\
\\
0.55 & -0.11 & -0.22 & -0.42 & -0.59 & -0.73 & -0.84 & -0.91 & -0.95 & -0.97 & -0.96 & -0.94 \\
0.60 & -0.13 & -0.25 & -0.49 & -0.69 & -0.86 & -0.99 & -1.08 & -1.13 & -1.15 & -1.15 & -1.13 \\
0.65 & -0.15 & -0.30 & -0.57 & -0.82 & -1.01 & -1.17 & -1.27 & -1.34 & -1.37 & -1.37 & -1.35 \\
0.70 & -0.18 & -0.35 & -0.67 & -0.96 & -1.20 & -1.38 & -1.51 & -1.59 & -1.63 & -1.64 & -1.62 \\
0.75 & -0.21 & -0.41 & -0.80 & -1.14 & -1.43 & -1.65 & -1.80 & -1.90 & -1.96 & -1.97 & -1.96 \\
0.80 & -0.25 & -0.50 & -0.97 & -1.38 & -1.73 & -2.00 & -2.19 & -2.32 & -2.39 & -2.41 & -2.40 \\
0.85 & -0.31 & -0.62 & -1.20 & -1.72 & -2.15 & -2.49 & -2.73 & -2.89 & -2.99 & -3.03 & -3.02 \\
0.90 & -0.41 & -0.82 & -1.58 & -2.26 & -2.83 & -3.27 & -3.60 & -3.82 & -3.96 & -4.02 & -4.02 \\
0.95 & -0.62 & -1.23 & -2.39 & -3.42 & -4.28 & -4.96 & -5.46 & -5.81 & -6.02 & -6.13 & -6.15 \\
0.99 & -1.46 & -2.90 & -5.62 & -8.05 & -10.1\,\,\, & -11.7\,\,\, & -12.9\,\,\, & -13.7\,\,\, & -14.2\,\,\, & -14.5\,\,\, & -14.6\,\,\, \\
\bottomrule
\end{tabular}
\end{table*}

For sample values of $x$ and $y$, the total radiative correction $\delta(x,y)$ is evaluated in Tab.~\ref{tab:deltaSym}.
The numbers here stem from the virtual and bremsstrahlung correction discussed in Sections~\ref{sec:virt} and \ref{sec:BS}, respectively.
Note that the correction to the $\Sigma^0\Lambda\gamma$ vertex $\delta_{\Sigma^0\Lambda\gamma}^\text{virt}(x,y)$ is not included here and will be discussed further.
The total radiative correction to the one-fold differential decay width $\delta(x)$ is shown in Fig.~\ref{fig:deltax}.
While the contribution of the virtual corrections is negative nearly throughout the whole region of electron--positron invariant mass, the bremsstrahlung is mainly positive (up to the region $x\gtrsim0.75$).
The overall correction is then above zero only for $x\lesssim0.25$, which is enough to make the correction to the decay rate positive: $\delta=0.896$; see also Tab.~\ref{tab:R}.
This is caused by the fact that the biggest contribution to the rate $\Gamma(\Sigma^0\to\Lambda e^+e^-)$ comes from the small-$x$ region of $\diff\Gamma(x)/\diff x$.

Integrating the differential decay width over the whole Dalitz plot and normalizing to the $\Sigma^0\to\Lambda\gamma$ decay rate, we can obtain the ratio \eqref{eq:R}.
Neglecting the effects of the electric form factor, we can write the result in terms of the linear expansion \eqref{eq:Rsplitup-magrad}.
At LO, taking into account the smallness of parameters $\rho$ and $\nu^2$, one can obtain simple analytic expressions which read
\begin{align}
R_0^\text{LO}
&\simeq\frac\alpha\pi
\left[
\frac23\log\left(\frac4\nu\right)
-\frac{13}9
-\frac\rho{15}
+\frac{\nu^2}4(3+\rho)
\right],\\
R_1^\text{LO}
&\simeq\frac\alpha\pi
\left[
\frac4{15}
-\nu^2
\right].
\end{align}
Note that the terms linear in $\rho$ and $\nu^2$ are numerically insignificant given the precision we use in Tab.~\ref{tab:R}.
Hence we can simply write
\begin{equation}
R^\text{LO}
\simeq\frac\alpha\pi\left[\frac23\log\left(\frac{2\Delta_M}{m}\right)-\frac{13}9+\frac4{15}a\right].
\end{equation}
Numerically, at NLO and taking into account $\delta^\text{virt}$ and $\delta^\text{BS}$ (i.e.\ consistently with corrections from Tab.~\ref{tab:deltaSym} or Fig.~\ref{fig:deltax}) reveals
$R
=5.544(2)\times10^{-3}$\,.
Neglecting the effects of the electric form factor, we can again express the result in terms of the linear expansion in $a$.
This is consistent with Ref.~\cite{Sidhu:1972rx} and allows us to isolate the form-factor effects:
\begin{equation}
R=(5.533+0.628a)\times10^{-3}\,.
\label{eq:Rour}
\end{equation}
This should be --- and perfectly is --- consistent with the NLO result for the rate given in Ref.~\cite{Sidhu:1972rx}, which was obtained using a different method:
\begin{equation}
R_\text{S\&S}
=(5.532+0.627a)\times10^{-3}\,.
\label{eq:RSS}
\end{equation}
We recall that $a$ is related to the slope of the magnetic form factor as $G_\text{M}(\Delta_M^2x)=G_\text{M}(0)(1+ax)$ or in other words, we have the relation \eqref{eq:defamagrad}.
On top of that, it is interesting to see how the respective corrections at NLO contribute to this result for $R$.
This is shown in Tab.~\ref{tab:R}.
Note that numerically $a=0.0183(26)$.

The situation changes only slightly if we take into account the QED corrections to the $\Sigma^0\Lambda\gamma$ vertex.
As discussed at the end of Section~\ref{sec:SLg}, its most significant contribution is related to the correction to the magnetic moment: $\delta_{\Sigma^0\Lambda\gamma}^\text{virt}\simeq2\delta_\kappa=-0.084(38)\,\%$.
This might be of a small effect when calculating the decay widths of the Dalitz decay $\Sigma^0\to\Lambda e^+e^-$ or of the radiative decay $\Sigma^0\to\Lambda\gamma$.
In this case, the previously specified correction can be used together with a theoretical prediction for the purely hadronic form-factor parameters.
Building the ratio $R$ or, equivalently, extracting the ``measured'' magnetic moment from the measurement of $\Sigma^0\to\Lambda\gamma$ and using it in $\Sigma^0\to\Lambda e^+e^-$ requires no need for this correction.
This is only necessary if one would like to disentangle the hadronic from the QED effects based on a future high-precision measurement.
For the measurement of the magnetic-form-factor slope, on the other hand, it is important to know how much the corrections to the $\Sigma^0\Lambda\gamma$ vertex affect the magnetic radius.
We find $\delta_{\langle r_\text{M}^2\rangle}=0.0071_{-26}^{+40}\,\%$, and thus they are very tiny.
Note that due to $\delta_{\langle r_\text{M}^2\rangle}$ the value for the ratio $R$ is modified only negligibly via its linear dependence on the corrected slope $a$.

Now, we also should take into account that we decided to completely neglect the contribution of the bremsstrahlung related to the baryon legs.
Moreover, we can also consider the unknown next-to-next-to-leading order (NNLO) QED correction as a source of additional uncertainty to our calculation.
After inspecting Tab.~\ref{tab:R}, we see that even though the overall NLO correction $\delta$ is below $1\,\%$, the contribution of $\text{BS}|_\text{D}$ itself is significantly bigger.
Suppose we assume that the NNLO correction could be of the same size with respect to NLO as $\delta^{\text{BS}|_\text{D}}$ is as compared to LO.
This means $\approx2.5\,\%$, and $\approx5\,\%$ for the slope.
We take the squares of these relative uncertainties to conservatively estimate an upper bound for the uncertainty of our final result:
\begin{equation}
R
=[5.533(3)+0.628(2)a]\times10^{-3}\,.
\end{equation}
Here, the uncertainty based on neglecting the higher-order calculations is bigger than the correction to the slope stemming from the QED corrections to the $\Sigma^0\Lambda\gamma$ vertex.

From the requirement that the branching ratios should sum up to 1,
\begin{equation}
\mathcal{B}(\Sigma^0\to\Lambda\gamma)
+\mathcal{B}(\Sigma^0\to\Lambda e^+e^-)
+\mathcal{B}(\Sigma^0\to\Lambda\gamma\gamma)
\simeq1\,,
\end{equation}
our knowledge of $R$ can be translated to the respective branching ratios.
We follow Ref.~\cite{Colas:1975ck} and assume $\mathcal{B}(\Sigma^0\to\Lambda\gamma\gamma)\approx10^{-7}$.
Then this part is irrelevant.
In view of our estimate on the uncertainty of $R$, we have
\begin{equation}
\mathcal{B}(\Sigma^0\to\Lambda\gamma)
\simeq\frac{1
}{1+R}\,,\quad
\mathcal{B}(\Sigma^0\to\Lambda e^+e^-)
\simeq\frac{R
}{1+R}\,.
\end{equation}
This becomes $\mathcal{B}(\Sigma^0\to\Lambda\gamma)=[99.4498(3)-0.0621(2)a]\,\%$, $\mathcal{B}(\Sigma^0\to\Lambda e^+e^-)=[0.5502(3)+0.0621(2)a]\,\%$.
Taking into account the value for the magnetic radius from Tab.~\ref{tab:num}, $\mathcal{B}(\Sigma^0\to\Lambda\gamma)=99.4486(5)\,\%$, $\mathcal{B}(\Sigma^0\to\Lambda e^+e^-)=0.5514(5)\,\%$.
Being even more conservative, let us assume 100\,\% uncertainty on the theoretical prediction of the magnetic radius of the $\Sigma^0\to\Lambda\gamma$ transition and consequently $a=0.02(2)$.
We find $\mathcal{B}(\Sigma^0\to\Lambda\gamma)=99.449(2)\,\%$, $\mathcal{B}(\Sigma^0\to\Lambda e^+e^-)=0.551(2)\,\%$.
These results are very reliable:
They are dominated by the QED calculation, assuming little about the size of the magnetic-form-factor slope, and can be further improved by future experimental knowledge of this parameter.

We now find ourselves in a position to summarize the content of this paper, emphasizing the differences with respect to Ref.~\cite{Sidhu:1972rx}.
That work was devoted to calculating both the corrections to the Dalitz plot and the virtual-photon spectrum, i.e.\ to the two-fold and one-fold differential decay widths, respectively, as well as to the decay rate.

In the former case, the soft-photon approximation was used together with an energy cut-off, which then comes in as a parameter when integrating over the degrees of freedom of the bremsstrahlung photon.
The presented expression is valid only for $x\gg\nu^2$.
It covers neither the hard-photon corrections nor the low-$x$ soft-photon corrections.
As pointed out by the authors of Ref.~\cite{Sidhu:1972rx}, due to this fact the resulting corrections were negative all over the Dalitz plot, in contrast to the fact that the total correction to the decay rate was found to be positive.
Indeed, the low-$x$ region is most important to correctly obtain the correction to the decay rate after the integral is performed.
In contrast, our calculation is valid over the whole Dalitz plot and includes the hard-photon corrections, simply because we performed the exact calculation without putting any extra limits on the photon energies.
Such {\em inclusive} radiative corrections are to be used in experiments when photons in the final state are ignored completely.

On the other hand, the correction to the decay rate obtained in Ref.~\cite{Sidhu:1972rx} contains the hard-photon corrections and that is why, when the corrections found in our work are integrated over the Dalitz plot, our result (\ref{eq:Rour}) is consistent with Eq.~(\ref{eq:RSS}).
This suits us as a neat cross-check of our calculation; the slight difference is caused by distinct inputs for masses and fine-structure constant.
Note that we also use muon loops as part of the virtual corrections and sum the whole geometric series of the vacuum-polarization insertions, though these effects tend to cancel each other to a very large extent.

To summarize, the (numerically) most significant difference between the previous work \cite{Sidhu:1972rx} and our present calculation is stemming from the way how the bremsstrahlung correction is treated in the case of the corrections to the differential decay width.
Yet, a second difference is that we also decided to explicitly calculate the contributions of additional loop diagrams which were omitted in Ref.~\cite{Sidhu:1972rx}: the two-photon-exchange (1$\gamma$IR) contribution and the correction to the $\Sigma^0\Lambda\gamma$ vertex.
What we found by our explicit calculations can be split up into general findings and numerical results.
In Ref.~\cite{Sidhu:1972rx} it was claimed that the radiative correction to the $\Sigma^0\Lambda\gamma$ vertex does not influence, by its nature, a determination of the {\em slope} of the form factor.
In contrast, we find that, in principle, both the normalization of the decay and the form factor slope (i.e.\ the radius) are influenced.
Numerically, however, as one can only see after the calculation is performed, all the effects are found to be very small.
Both the $\Sigma^0\Lambda\gamma$-vertex correction and the 1$\gamma$IR contribution can thus be neglected in the evaluation of the NLO radiative corrections.

Finally, from Fig.~\ref{fig:deltax} we can estimate the size of the correction to the (magnetic) form-factor slope.
By taking half of the slope of the curve in the low-$x$ region, however farther from the threshold:
\begin{equation}
\Delta a
\equiv a_\text{(+QED)}-a
\simeq\frac12\frac{\diff\delta(x)}{\diff x}\bigg|_{x=x_0}\,,
\end{equation}
with $\nu^2\ll x_0\ll1$.
Since $\frac12\frac{\diff\delta(x)}{\diff x}\big|_{x=x_0}\approx-3.5\,\%$, this correction is bigger than the estimate on $a$ itself ($a\approx1.8(3)\,\%$).
What would this imply if one tried to extract a magnetic radius from experiment without considering the radiative corrections explicitly?
One would obtain a measured value of $a_\text{(+QED)}$ that implicitly contains the QED radiative corrections.
Thus one should then {\em subtract} from the measured value $a_\text{(+QED)}$ the correction $\Delta a$ in order to get an estimate on the hadronic parameter $a$, i.e.\ on the magnetic radius on account of Eq.~\eqref{eq:defamagrad}.
With the above assumed values, one would thus expect the ``measured'' radius to be negative, in case the radiative corrections would not be used explicitly in the analysis: $\langle r_\text{M}^2\rangle_\text{(+QED)}=\langle r_\text{M}^2\rangle+\frac6{\Delta_M^2}\Delta a$, with $\frac6{\Delta_M^2}\Delta a\approx-35\,\text{GeV}^{-2}$.

\begin{acknowledgements}
We want to thank P.\ S\'anchez Puertas for bringing Ref.~\cite{Fael:2016yle} to our attention.
This work has been supported in part by
Grants No.\ FPA2017-84445-P and
SEV-2014-0398 (AEI/ERDF, EU)
and by PROMETEO/2017/053 (GV).
\end{acknowledgements}

\appendix

\section{Bremsstrahlung: basic integrals}
\label{app:J}

For the purpose of this appendix, it is useful to define the following function:
\begin{equation}
L(a,b,c)
\equiv\frac1{\sqrt{a^2-b}}\log\Bigg|\frac{c+a+\sqrt{a^2-b}}{c+a-\sqrt{a^2-b}}\Bigg|\,.
\label{eq:Lab}
\end{equation}
It is also convenient to write explicitly the energies of the leptons in the Lambda--$\gamma$ CMS in the invariant form,
\begin{align}
\omega q_{1,0}&\stackrel{(\vec{r}=0)}{=}(k+p_2)\cdot q_1=\frac14\left(M_\Sigma^2-s-s_\gamma-\Delta_m^2\right),\\
\omega q_{2,0}&\stackrel{(\vec{r}=0)}{=}(k+p_2)\cdot q_2=\frac14\left(M_\Sigma^2-s-s_\gamma+\Delta_m^2\right);
\end{align}
see the text after Eq.~\eqref{eq:MBSconv} for the relation between the dimensionless variables $x$ and $y$ and the variables $s$ and $\Delta_m^2$.
The first set of basic integrals can be written as
\begin{gather}
J[1]
=\frac{\tilde\omega}\omega\,,\\
J\big[A\big]
=\frac{{\tilde\omega}^2}{\omega^2}\frac{\omega q_{2,0}}2\,,\\
J\bigg[\frac 1{A}\bigg]
=L(\omega q_{2,0},\omega^2m^2,0)\,,\\
J\bigg[\frac 1{E}\bigg]
=\frac12L\bigg(\omega (q_{1,0}+q_{2,0}),\omega^2s,\frac\omega{\tilde\omega} \tilde s\bigg)\,,\\
J\bigg[\frac 1{AB}\bigg]
=\frac8{{\tilde\omega}\omega}L(s,4m^2s,0)\,,\\
J\bigg[\frac 1{A^2}\bigg]
=\frac4{m^2{\tilde\omega}\omega}\,,\\
J\bigg[\frac 1{E^2}\bigg]
=\frac1s\frac{{\tilde\omega}\omega}{[{\tilde\omega}\omega(M_\Sigma^2-M_\Lambda^2)+sM_\Lambda^2]}\,.
\end{gather}

For the other set, it is useful to introduce additional variables,
\begin{gather}
v_1=\omega q_{2,0}\frac{\tilde\omega}\omega\,,\\
v_2=\omega q_{1,0}\frac{\tilde\omega}\omega+\tilde s\,,\\
w_0=m^2{\tilde\omega}^2\,,\\
w_1=(s-2m^2){\tilde\omega}^2+2\tilde s\omega q_{2,0}\frac{\tilde\omega}\omega\,,\\
w_2=m^2{\tilde\omega}^2+\tilde s^2+2\tilde s\omega q_{1,0}\frac{\tilde\omega}\omega\,,
\end{gather}
and their following combinations:
\begin{gather}
\varrho=2w_0+w_1\,,\\
\tilde\varrho=2w_2+w_1\,,\\
\varsigma=w_1^2-4w_0w_2\,,\\
w=w_0+w_1+w_2\,,\\
\tau_1=v_2\varrho-v_1\tilde\varrho\,,\\
\tau_2=v_1w_1-2v_2w_0\,,\\
\tau=v_1^2w_2+v_2^2w_0\,.
\end{gather}
The remaining basic integrals are then given as
\begin{align}
\begin{split}
J\bigg[\frac 1{AE}\bigg]
&=\frac{2{\tilde\omega}}\omega L(2w_0+w_1,4w_0(w_0+w_1+w_2),0)\\
&=\frac{2{\tilde\omega}}\omega\frac1{\sqrt{\varsigma}}\log\bigg[\frac{\varrho+\sqrt{\varsigma}}{\varrho-\sqrt{\varsigma}}\bigg]\,,
\end{split}\\
J\bigg[\frac 1{AE^2}\bigg]
&=\frac2\varsigma\left\{\tau_1\frac{2{\tilde\omega}}\omega\frac1w+\tau_2J\bigg[\frac 1{AE}\bigg]\right\},\\
J\bigg[\frac 1{A^2E}\bigg]
&=\frac4\varsigma\left\{\tau_2\frac{2{\tilde\omega}}\omega\frac1{w_0}+\tau_1J\bigg[\frac 1{AE}\bigg]\right\},\label{eq:JA2E}\\
\begin{split}
J\bigg[\frac 1{A^2E^2}\bigg]\\
&\hspace{-12mm}=\frac{16}\varsigma
\left\{
\frac{2{\tilde\omega}}{\omega}
\left[
\frac12\left(1+\frac{v_1^2}{w_0}+\frac{(v_1+v_2)^2}w\right)+\frac6\varsigma\left(\tau-v_1v_2w_1\right)
\right]\right.\\
&\hspace{-12mm}-\left.\left[
\frac\varrho4-2v_1v_2+v_1^2+\frac3\varsigma\left(\tau\varrho-2v_1v_2w_0\tilde\varrho\right)
\right]
J\bigg[\frac 1{AE}\bigg]
\right\},
\end{split}
\end{align}
and
\begin{equation}
J\bigg[\frac 1{X(E-4\hat M^2)}\bigg]
=J\bigg[\frac 1{XE}\bigg]_{\tilde s\to s-4\hat M^2}\,,
\end{equation}
with $X\in\{1,A,A^2\}$.
In the previous expressions, whenever $\tilde s$ is not substituted, in the end we put $\tilde s\to s$.
In the limit $M_\Lambda\to0$ (and consequently ${\tilde\omega}\to\omega$) and $M_\Sigma\to M_{\pi^0}$, the results from Ref.~\cite{Husek:2015sma} are recovered.
Note that $J\big[\frac 1{A^2E}\big]$, $J\big[\frac 1{A^2(E-4\hat M^2)}\big]$ and $J\big[\frac 1{A^2E^2}\big]$ contain divergent parts $\frac1sJ\big[\frac 1{A^2}\big]$, $\frac1{s-4\hat M^2}J\big[\frac 1{A^2}\big]$ and $\frac1{s^2}J\big[\frac 1{A^2}\big]$, respectively:
For instance, in Eq.~(\ref{eq:JA2E}), $\frac{\tilde\omega}\omega\frac4{w_0}=J\big[\frac 1{A^2}\big]$ and $\left(\frac{2\tau_2}\varsigma-\frac1s\right)J\big[\frac 1{A^2}\big]$ then leads to a convergent integral.
Needless to mention, the divergences have to be subtracted before the numerical integration is performed and treated separately:
The corresponding divergent integrals have to be integrated over $s_\gamma$ analytically.
The results can be written in a simple form:
\begin{align}
\begin{split}
&\int J\bigg[\frac 1{A^2}\bigg]\diff s_\gamma\\
&=\frac{4}{m^2}\bigg[\log\frac m \Lambda+\log\frac{2(s_\gamma^\text{max}-M_\Lambda^2)}{M_\Sigma^2-M_\Lambda^2-s+\Delta_m^2}\bigg]\,,
\end{split}
\label{eq:JA2div}\\
\begin{split}
&\int J\bigg[\frac 1{AB}\bigg]\diff s_\gamma
=\frac4{s\beta}\biggl\{-2\log(\gamma)\biggr.\\
&\times\bigg[\log\frac{m}{\Lambda}+\log\frac{2(s_\gamma^\text{max}-M_\Lambda^2)}{M_\Sigma^2-M_\Lambda^2-s}-\frac12\log(1-\beta^2)\bigg]\\
&+\left.K\bigg(\beta,\frac{\Delta_m^2}{M_\Sigma^2-M_\Lambda^2-s}\bigg)\right\},
\end{split}
\label{eq:JABdiv}
\end{align}
where
\begin{equation}
\begin{split}
K(\beta,y)
&=2\log(\gamma)\log\bigg(\frac{y+\beta}{2\beta}\bigg)\\
&-\text{Li}_2\bigg[\frac{\gamma(y-\beta)}{y+\beta}\bigg]+\text{Li}_2\bigg[\frac{y-\beta}{\gamma(y+\beta)}\bigg]\,,
\end{split}
\label{eq:Kxy}
\end{equation}
and with
\begin{equation}
s_\gamma^\text{max}
=M_\Sigma^2+s-\sqrt{4sM_\Sigma^2+\frac1{\beta^2}(\Delta_m^2)^2}\,.
\end{equation}

To list the last missing basic integrals, we define the following variables:
\begin{align}
b_i&=\frac{q_{i,0}}{|\vec q_i|}\,,\quad
|\vec q_i|=\sqrt{q_{i,0}^2-m^2}\,,\\
\eta&=\frac{\vec q_1\cdot\vec q_2}{|\vec q_1||\vec q_2|}
=b_1b_2-\frac12\frac{(s-2m^2)}{|\vec q_1||\vec q_2|}\,,
\end{align}
and functions:
\begin{align}
\tilde Q_1(\xi)
&\equiv b_1Q_0(\xi)-\eta Q_1(\xi)\,,\\
\begin{split}
\tilde Q_2(\xi)
&\equiv\bigg(\frac13+b_1^2\bigg)Q_0(\xi)\\
&-2b_1\eta Q_1(\xi)+\bigg(\eta^2-\frac13\bigg)Q_2(\xi)\,,
\end{split}
\end{align}
where $Q_m(\xi)$ are the Legendre functions of second kind, with $Q_0(\xi)$ redefined for $\xi>1$:
\begin{align}
Q_0(\xi)
&=\frac12\log\bigg(\frac{\xi+1}{\xi-1}\bigg)\,,\\
Q_1(\xi)
&=\xi Q_0(\xi)-1\,,\\
Q_2(\xi)
&=\frac12(3\xi^2-1)Q_0(\xi)-\frac32\xi\,.
\end{align}
The integrals then read
\begin{align}
J\bigg[\frac BA\bigg]
&=\frac{\tilde\omega}\omega\frac{|\vec q_1|}{|\vec q_2|}\tilde Q_1(b_2)\,,\\
J\bigg[\frac B{A^2}\bigg]
&=-\frac2\omega\frac{|\vec q_1|}{|\vec q_2|^2}\frac{\diff\tilde Q_1(\xi)}{\diff\xi}\bigg|_{\xi=b_2}\,,\\
J\bigg[\frac {B^2}A\bigg]
&=\frac{{\tilde\omega}^2}{2\omega}\frac{|\vec q_1|^2}{|\vec q_2|}\tilde Q_2(b_2)\,,\\
J\bigg[\frac {B^2}{A^2}\bigg]
&=-\frac{\tilde\omega}\omega\frac{|\vec q_1|^2}{|\vec q_2|^2}\frac{\diff\tilde Q_2(\xi)}{\diff\xi}\bigg|_{\xi=b_2}\,.
\end{align}

\section{\texorpdfstring{1$\gamma$IR}{1gIR} contribution: the baryonic part}
\label{app:betaT}

For further convenience, let us define the product of two form factors appearing in the 1$\gamma$IR amplitude (see Eq.~(\ref{eq:Btilde})):
\begin{equation}
g_{ij}^X=G_i^{X\Lambda}\big((l-p_2)^2\big)\,G_j^{\Sigma^0X}\big((l-p_1)^2\big)\,.
\end{equation}
then the coefficients and matrices in Eq.~(\ref{eq:BX}) take the form
\begin{alignat}{2}
T_{\mu\nu}^1&\equiv l^\rho\gamma_{\mu\rho\nu}\,,
&&\beta_1^X\equiv g_{22}^X\,\frac{l^2-M_X^2}{2M_X(M_\Sigma+M_\Lambda)}+\sum_{i,j=1}^2g_{ij}^X\,,\\
T_{\mu\nu}^2&\equiv\gamma_{\mu\nu}\,,
&&\beta_2^X\equiv M_X\beta_1^X+(l^2-M_X^2)\notag\\
& &&\times
\bigg(\frac{g_{12}^X}{M_X+M_\Sigma}+\frac{g_{21}^X}{M_X+M_\Lambda}+\frac{g_{22}^X}{2M_X}\bigg)\,,\\
T_{\mu\nu}^3&\equiv2l_\nu l^\rho\gamma_{\mu\rho}\,,
&&\beta_3^X\equiv-\frac{g_{22}^X+g_{12}^X}{M_X+M_\Sigma}\,,\\
T_{\mu\nu}^4&\equiv2l_\mu l^\rho\gamma_{\rho\nu}\,,
&&\beta_4^X\equiv-\frac{g_{22}^X+g_{21}^X}{M_X+M_\Lambda}\,,\\
T_{\mu\nu}^5&\equiv2l_\nu\gamma_\mu\,,
&&\beta_5^X\equiv M_X\beta_3^X-g_{22}^X\,\frac{l^2-M_X^2}{2M_X(M_\Sigma+M_\Lambda)}\,,\\
T_{\mu\nu}^6&\equiv2l_\mu\gamma_\nu\,,
&&\beta_6^X\equiv M_X\beta_4^X-g_{22}^X\,\frac{l^2-M_X^2}{2M_X(M_\Sigma+M_\Lambda)}\,,\\
T_{\mu\nu}^7&\equiv4l_\mu l_\nu l^\rho\gamma_\rho\,,\;\,
&&\beta_7^X\equiv g_{22}^X\,\frac{1}{2M_X(M_\Sigma+M_\Lambda)}\,,\\
T_{\mu\nu}^8&\equiv4l_\mu l_\nu\,,
&&\beta_8^X\equiv M_X\beta_7^X\,,
\end{alignat}
where we used for simplicity the short-hand notation \eqref{eq:gamma_prod} for a product of $\gamma$-matrices.
We also extracted the loop momenta out of the expressions.
Now, due to the conservation law of the electromagnetic current (\ref{eq:L_gauge}), the loop momenta with the Lorentz indices $\mu$ and $\nu$ can be substituted in the following manner:
\begin{equation}
l^\mu\to p_2^\mu\,,\quad l^\nu\to p_1^\nu\,.
\end{equation}
Loop momenta carrying index $\rho$ then enter the tensorial one-loop integrals defined in \ref{app:LI}.

\section{A model for the \texorpdfstring{$\Sigma^0\to\Lambda\gamma$}{S->Lg} transition form factor}
\label{app:FF}

Having a particular model for the form factors is essential in the case of the 1$\gamma$IR contribution and the correction to the $\Sigma^0\Lambda\gamma$ vertex.
There the form factors enter loop integrals and therefore a low-energy expansion like in Eq.~\eqref{eq:expandGEGM} is insufficient.

In the context of the whole box diagram in the case of the 1$\gamma$IR contribution, the terms proportional to $G_2(q^2)$ are potentially responsible for the UV-divergent behavior due to the loop-momenta-power counting; cf.\ Eq.~(\ref{eq:Btilde}).
Following Eq.~(\ref{eq:GEM}) we get
\begin{equation}
\begin{split}
G_1^{XY}(q^2)
&=\frac{q^2G_\text{M}^{XY}(q^2)-(M_X+M_Y)^2G_\text{E}^{XY}(q^2)}{q^2-(M_X+M_Y)^2}\,,\\
G_2^{XY}(q^2)
&=\frac{(M_X+M_Y)^2\big(G_\text{E}^{XY}(q^2)-G_\text{M}^{XY}(q^2)\big)}{q^2-(M_X+M_Y)^2}\,,
\end{split}
\label{eq:G12b}
\end{equation}
and we see that for $q^2\to\infty$ and {\em constant} $G_\text{E}^{XY}(q^2)$ and $G_\text{M}^{XY}(q^2)$, $G_1^{XY}(q^2)\simeq G_\text{M}^{XY}(q^2)$ and $|G_2^{XY}(q^2)|\sim1/q^2$.
Hence, regarding the 1$\gamma$IR contribution, the UV convergence is achieved even in this simplest case when constant form factors are assumed.
Taking $G_\text{E}^{XY}(q^2)=G_\text{E}^{XY}(0)=0$ and $G_\text{M}^{XY}(q^2)=G_\text{M}^{XY}(0)=\kappa^{XY}$, one arrives at
\begin{equation}
\begin{split}
G_1^{XY}(q^2)&=\kappa^{XY}\frac{q^2}{q^2-M_V^2}\,,\\
G_2^{XY}(q^2)&=-\kappa^{XY}\frac{M_V^2}{q^2-M_V^2}\,.
\end{split}
\label{eq:GEM_const}
\end{equation}
It could be then sufficient to show that if one uses this simple prescription, then the 1$\gamma$IR contribution is negligible.
Such a conclusion should then carry over to the cases of more sophisticated models.

To explore also a second possibility, we use an ansatz exhibiting a stronger suppression in the UV region of the form
\begin{equation}
\begin{split}
G_1^{XY}(q^2)&=c_1^{XY}\frac{q^2M_V^4}{(q^2-M_V^2)^3}\,,\\
G_2^{XY}(q^2)&=c_2^{XY}\frac{M_V^6}{(q^2-M_V^2)^3}\,,
\end{split}
\label{eq:GEM_nonconst}
\end{equation}
which satisfies the Brodsky--Lepage scaling rules~\cite{Lepage:1979za,Lepage:1980fj} and --- for nontrivial $c_1^{XY}$ and $c_2^{XY}$ --- the conditions $G_\text{E}^{XY}(0)=0$ and $G_\text{M}^{XY}(0)=\kappa^{XY}$.
After inserting these expressions in Eq.~(\ref{eq:GEM}) we find $c_2^{XY}=-\kappa^{XY}$ and
\begin{align}
c_{1,\text{E}}^{XY}&=\kappa^{XY}\frac{M_V^2}{(M_X+M_Y)^2}-\frac16\langle r_\text{E}^2\rangle^{XY} M_V^2\,,\label{eq:c1E}\\
c_{1,\text{M}}^{XY}&=\kappa^{XY}\left(3-\frac16\langle r_\text{M}^2\rangle^{XY} M_V^2\right)\label{eq:c1M},
\end{align}
depending on if we match the linear expansions in $q^2$ to $G_\text{E}^{XY}(q^2)$ or $G_\text{M}^{XY}(q^2)$; see Eqs.~\eqref{eq:expandGEGM}.
In the present work we take $c_1^{XY}\equiv c_{1,\text{M}}^{XY}$.

\section{Loop structure in the \texorpdfstring{1$\gamma$IR}{1gIR} contribution}
\label{app:BL}

In this appendix we write the contribution of the box diagram to the matrix element squared at NLO in terms of tensorial one-loop integrals, taking into account the model introduced in \ref{app:FF}.
The terms having a nontrivial loop-momentum dependence, listed by the source, are the leptonic part contributing with
\begin{equation}
\frac{\{l^\kappa\}}{(l-p_2-q_1)^2-m^2}\,,
\end{equation}
the baryonic part giving
\begin{equation}
\frac{\{l^\rho\}\{l^2-M_X^2\}}{l^2-M_X^2}\,,
\label{eq:barC}
\end{equation}
photon propagators
\begin{equation}
\frac1{(l-p_1)^2(l-p_2)^2}\,,
\end{equation}
and form factors
\begin{equation}
\frac{\{(l-p_1)^2\}\{(l-p_2)^2\}}{[(l-p_1)^2-M_{(X\Lambda)}^2]^n[(l-p_2)^2\{-M_{(\Sigma^0X)}^2\}]^n}\,.
\end{equation}
Note that terms in curly brackets appear only in part of the terms.
This means we have tensorial integrals of the triangle ($C$) and box ($D$) type.
The rank of these tensors ranges up to 2.

In what follows, we want to regroup the terms listed in \ref{app:betaT} based on $g_{ij}^X$ and associate them with the tensorial integrals of a given type.
It is convenient to introduce the loop-momenta-independent traces
\begin{align}
\text{Tr}_{\mu\{\rho\}\nu,\sigma}^{\text{B}(11),X}
&\equiv
\text{Tr}_{\mu\{\rho\}\nu,\sigma}^{\text{B}}\,,\\
\text{Tr}_{\mu\{\rho\}\nu,\sigma}^{\text{B}(12),X}
&\equiv
\text{Tr}_{\mu\{\rho\}\nu,\sigma}^{\text{B}}
-\frac{2p_{1\nu}}{M_X+M_\Sigma}\text{Tr}_{\mu\{\rho\},\sigma}^{\text{B}}\,,\\
\text{Tr}_{\mu\{\rho\}\nu,\sigma}^{\text{B}(21),X}
&\equiv
\text{Tr}_{\mu\{\rho\}\nu,\sigma}^{\text{B}}
-\frac{2p_{2\mu}}{M_X+M_\Lambda}\text{Tr}_{\{\rho\}\nu,\sigma}^{\text{B}}\,,\\
\begin{split}
\text{Tr}_{\mu\{\rho\}\nu,\sigma}^{\text{B}(22),X}
&\equiv
\text{Tr}_{\mu\{\rho\}\nu,\sigma}^{\text{B}}
+\frac{4p_{2\mu}p_{1\nu}}{2M_X(M_\Sigma+M_\Lambda)}\text{Tr}_{\{\rho\},\sigma}^{\text{B}}\\
&-\frac{2p_{1\nu}}{M_X+M_\Sigma}\text{Tr}_{\mu\{\rho\},\sigma}^{\text{B}}
-\frac{2p_{2\mu}}{M_X+M_\Lambda}\text{Tr}_{\{\rho\}\nu,\sigma}^{\text{B}}\,,
\end{split}
\end{align}
which then combine (so far with general masses $M_1$ and $M_2$) with the box integrals as
\begin{equation}
\begin{split}
\text{DTr}_{\{\kappa\},\mu\nu,\sigma}^{\text{B}(ij),X}(M_1^2,M_2^2)
&\equiv
D_{\{\kappa\}}^{X\rho}(M_1^2,M_2^2)\text{Tr}_{\mu\rho\nu,\sigma}^{\text{B}(ij),X}\\
&+D_{\{\kappa\}}^X(M_1^2,M_2^2)\text{Tr}_{\mu\nu,\sigma}^{\text{B}(ij),X}M_X\,.
\end{split}
\end{equation}
Similarly, let us define
\begin{align}
&\text{CTr}_{\{\kappa\},\mu\nu,\sigma}^{\text{B}(11),X}(M_1^2,M_2^2)
\equiv0\,,\\
&\text{CTr}_{\{\kappa\},\mu\nu,\sigma}^{\text{B}(12),X}(M_1^2,M_2^2)
\equiv
\frac1{M_X+M_\Sigma}\,
C_{\{\kappa\}}(M_1^2,M_2^2)\text{Tr}_{\mu\nu,\sigma}^{\text{B}}\,,\\
&\text{CTr}_{\{\kappa\},\mu\nu,\sigma}^{\text{B}(21),X}(M_1^2,M_2^2)
\equiv
\frac1{M_X+M_\Lambda}\,
C_{\{\kappa\}}(M_1^2,M_2^2)\text{Tr}_{\mu\nu,\sigma}^{\text{B}}\,,\\
\begin{split}
&\text{CTr}_{\{\kappa\},\mu\nu,\sigma}^{\text{B}(22),X}(M_1^2,M_2^2)
\equiv
\frac1{2M_X(M_\Sigma+M_\Lambda)}\\
&\times\biggl\{C_{\{\kappa\}}^{\rho}(M_1^2,M_2^2)\text{Tr}_{\mu\rho\nu,\sigma}^{\text{B}}
+C_{\{\kappa\}}(M_1^2,M_2^2)\biggr.\\
&\times\biggl.\big[(M_\Sigma+M_\Lambda+M_X)\text{Tr}_{\mu\nu,\sigma}^{\text{B}}
-2p_{1\nu}\text{Tr}_{\mu,\sigma}^{\text{B}}
-2p_{2\mu}\text{Tr}_{\nu,\sigma}^{\text{B}}\big]
\biggr\}
\end{split}
\end{align}
to gain the triangle-type contributions arising due to the cancellation of one of the denominators, as shown in~(\ref{eq:barC}).
The overall generic sum of the products of the tensorial loop integrals and traces of the baryonic part is then
\begin{equation}
\text{LTr}_{\{\kappa\},\mu\nu,\sigma}^{\text{B}(ij),X}
\equiv
\text{CTr}_{\{\kappa\},\mu\nu,\sigma}^{\text{B}(ij),X}
+\text{DTr}_{\{\kappa\},\mu\nu,\sigma}^{\text{B}(ij),X}\,.
\end{equation}
A subsequent contraction with the leptonic part (in terms of Eq.~(\ref{eq:Labt})) then yields
\begin{equation}
\begin{split}
&\text{Tr}_{\sigma\tau}^{\text{BL}(ij),X}(M_1^2,M_2^2;q_1,q_2)\\
&\equiv
\text{LTr}_{,\mu\nu,\sigma}^{\text{B}(ij),X}(M_1^2,M_2^2)
\text{Tr}_{\quad\;\tau}^{\text{L}\mu\nu}(q_1,q_2)\\
&-\text{LTr}_{\kappa,\mu\nu,\sigma}^{\text{B}(ij),X}(M_1^2,M_2^2)
\text{Tr}_{\quad\;\;\;\tau}^{\text{L}\mu\kappa\nu}(q_1,q_2)\,,
\end{split}
\label{eq:TrBLijX}
\end{equation}
which serves as the desired building block.

\section{Tensorial one-loop integrals}
\label{app:LI}

We use the tensorial one-loop integrals as defined below.
In what follows we use for Feynman denominators $D_0\equiv[l^2-m_0^2+i\epsilon]$ and $D_i\equiv D(p_i,m_i^2)\equiv[(l-p_i)^2-m_i^2+i\epsilon]$ for $i\ge1$.
\begin{align}
&B_{\kappa..\rho}\big(p_1^2;m_0^2,m_1^2\big)
\equiv\int\hskip-1mm\frac{\diff^4l}{(2\pi)^4}\,
\frac{l_\kappa..\,l_\rho}{D_0D_1}\,,\\
&C_{\kappa..\rho}\big(p_1^2,(p_2-p_1)^2,p_2^2;m_0^2,m_1^2,m_2^2\big)
\equiv\int\hskip-1mm\frac{\diff^4l}{(2\pi)^4}\,
\frac{l_\kappa..\,l_\rho}{D_0D_1D_2}\,,\\
\begin{split}
&D_{\kappa..\rho}\big(p_1^2,(p_2-p_1)^2,(p_3-p_2)^2,p_3^2,p_2^2,(p_3-p_1)^2;\\
&\hspace{15mm}m_0^2,m_1^2,m_2^2,m_3^2\big)
\equiv\int\hskip-1mm\frac{\diff^4l}{(2\pi)^4}\,
\frac{l_\kappa..\,l_\rho}{D_0D_1D_2D_3}\,.
\end{split}
\end{align}
During the calculation of the 1$\gamma$IR contribution stemming from the box diagrams, it is only necessary to take into account the integrals $C_{\kappa..\rho}(M_{(X\Lambda)}^2,M_{(\Sigma^0X)}^2)$ and $D_{\kappa..\rho}^X(M_{(X\Lambda)}^2,M_{(\Sigma^0X)}^2)$ defined as
\begin{align}
\begin{split}
&C_{\kappa..\rho}(M_{(X\Lambda)}^2,M_{(\Sigma^0X)}^2)\\
&\equiv C_{\kappa..\rho}\big((q_1+q_2)^2,m^2,m^2;M_{(\Sigma^0X)}^2,M_{(X\Lambda)}^2,m^2\big)\\
&=\int\hskip-1mm\frac{\diff^4l}{(2\pi)^4}\,\frac{l_\kappa..\,l_\rho}
{D(p_1,M_{(\Sigma^0X)}^2)D(p_2,M_{(X\Lambda)}^2)D(p_2+q_1,m^2)}\,,
\end{split}\\
\begin{split}
&D_{\kappa..\rho}^X(M_{(X\Lambda)}^2,M_{(\Sigma^0X)}^2)\\
&\equiv D_{\kappa..\rho}\big(M_\Sigma^2,(q_1+q_2)^2,m^2,(p_2+q_1)^2,M_\Lambda^2,m^2;\\
&\hspace{15mm}M_X^2,M_{(\Sigma^0X)}^2,M_{(X\Lambda)}^2,m^2\big)\\
&=\int\hskip-1mm\frac{\diff^4l}{(2\pi)^4}\,\bigg\{\frac1
{D(0,M_X^2)}\\
&\hspace{15mm}\times\frac{l_\kappa..\,l_\rho}{D(p_1,M_{(\Sigma^0X)}^2)D(p_2,M_{(X\Lambda)}^2)D(p_2+q_1,m^2)}\bigg\}\,.
\end{split}
\end{align}


\providecommand{\href}[2]{#2}
\end{document}